\documentstyle[aaspp4]{article}

\def\go{\mathrel{\raise.3ex\hbox{$>$}\mkern-14mu\lower0.6ex\hbox{$\sim$}}}
\def\lo{\mathrel{\raise.3ex\hbox{$<$}\mkern-14mu\lower0.6ex\hbox{$\sim$}}}

\begin{document}

\title{Structure and Evolution of Nearby Stars with Planets. \\
I. Short-Period Systems}
\author{Eric B.\ Ford, Frederic A.\ Rasio}

\affil{Department of Physics, MIT 6-201, Cambridge, MA 02139\\
eford@mit.edu; rasio@mit.edu}
\and
\author {Alison Sills\altaffilmark{1}}
\affil{Department of Astronomy, Yale University, P.O.\ Box 208101,
New Haven, CT 06520}
\altaffiltext{1}{Present Address:  Ohio State University; asills@astronomy.ohio-state.edu}
\authoremail{eford@mit.edu}

\begin{abstract}

Using the Yale stellar evolution code, we have calculated
theoretical models for nearby stars with planetary-mass companions in
short-period nearly circular orbits: 51 Pegasi, \hbox{$\tau$ Bootis}, $\upsilon$
Andromedae, $\rho^1$ Cancri, and $\rho$ Coronae Borealis.  We present
tables listing key stellar parameters such as mass, radius, age, and
size of the convective envelope as a function of the observable
parameters (luminosity, effective temperature, and metallicity), as
well as the unknown helium fraction.  For each star we construct
best models based on recently published spectroscopic data and the
present understanding of galactic chemical evolution.  We discuss our
results in the context of planet formation theory, and, in particular,
tidal dissipation effects and stellar metallicity enhancements.

\end{abstract}


\section{Introduction and Motivation}

The detection of planets outside the Solar System constitutes one of
the most exciting recent developments in astronomy and astrophysics.
These discoveries of extrasolar planets will lead to significant
improvements in our understanding of many processes related to planet
and star formation, as well as deeper questions such as the existence
of extraterrestrial life in the Universe.  We now know of more planets
outside the Solar System than inside (including planets around
pulsars; Wolszczan 1994).  Several groups have reported detections of
Jupiter-type planets around nearby solar-like stars (see Table 1), and
new announcements continue to arrive every few months.  We expect that
radial velocity surveys already in progress (Marcy et al.~1997;
Korzennik et al.~1997; Mayor \& Queloz 1997; Cochran et al.~1997a;
Butler et al.~1998) will discover many more in the near future.  Other
techniques such as photometry (Borucki \& Summers 1984) and astrometry
(Gatewood 1996; Pravdo \& Shaklan 1996), including interferometric
astrometry (Colavita \& Shao 1994) and, in the future,
space-based interferometric astrometry (Unwin et al.~1996; Boden et
al.~1996), may lead to additional detections.  In addition,
observations of protostellar disks (Mannings 1998) and T Tauri stars
(Beckwith \& Sargent 1996) can provide important information about
planet formation.  In the long term, space based interferometry
appears very promising (Unwin et al.~1996; Boden et al.~1996).

This paper is the first of a series in which we use the
Yale Rotating Evolution Code (the YREC code developed
by Demarque and collaborators; see Guenther et al.\ 1992) to
calculate theoretical models for the present structure and past
evolution of stars known to harbor planets, using recently
measured stellar parameters and Hipparcos parallaxes as
constraints. Our goal is to obtain a set of self-consistent
theoretical models that will provide the best possible values (with
realistic error estimates) for all derived stellar parameters, such
as mass and radius, which are crucial for a number of theoretical
investigations of these systems.  In this first paper we present a general
description of the stellar evolution code and our method for
constructing the theoretical models. 

\subsection{Tidal Dissipation} 

Our models will provide a detailed description of the stellar
convective zone, which is essential when calculating tidal dissipation
through eddy viscosity (e.g., Zahn 1977; Zahn \& Bouchet 1989).  Tidal
dissipation may have played a crucial role in the formation and
evolution of the 51 Peg-type systems. Despite observational biases,
the clustering of orbital periods around $3-4\,$ d for these systems
is probably not a coincidence.  The San Francisco State University
(SFSU) extrasolar planet search (Marcy et al.~1997) is capable of
detecting periodic radial velocity variations with an amplitude of 12
m s$^{-1}$, corresponding to a 1$M_{J}$ planet orbiting at 1 AU from a
1 $M_{\odot}$ star given a favorable inclination.  Additional factors,
such as the variability of stellar photospheres and the longer time
span of observations necessary to confirm the Keplerian nature of the
radial velocity variations, can also inhibit detection of planets with
larger semi-major axes. Nevertheless, the present low yield of
extrasolar planets with larger semi-major axes is likely indicative of
a real pattern.  Periods of $3-4\,$ d correspond to the onset of
strong tidal dissipation in such systems. When the star is rapidly
rotating, tidal torques may prevent the planet from spiraling all the
way into the star and being destroyed (Lin, Bodenheimer, \& Richardson
1996; Trilling et al.\ 1998). This may be essential in formation
scenarios invoking slow inward migration of giant planets in a
protostellar disk. Alternatively, tidal dissipation effects may
circularize a highly eccentric orbit produced by a dynamical
interaction (Rasio \& Ford 1996).  In addition, for a slowly rotating
star (most of the stars in Table~1 are known to be slowly rotating
today), tidal dissipation leads to (possibly rapid) {\it orbital
decay\/}, and therefore the survival of the system to the present
provides an important constraint on theoretical models (Rasio et
al.~1996).  In some cases this may also provide an upper limit on the
companion mass (see \S 4.1).

Tidal dissipation is particularly interesting yet difficult to study
theoretically in these systems since the tidal pumping period (half
the orbital period) is short compared to the typical eddy turnover
timescale in the stellar convective envelope. In this regime, the
efficiency of tidal dissipation can be drastically reduced (Goldreich
\& Keeley 1977), although the details of the theory in this case are
rather controversial (see Goodman \& Oh 1997 and references therein).
The 51 Peg-type systems therefore provide important new constraints on
the theory of tidal dissipation in a regime where it is particularly
poorly understood. A better understanding of tidal effects (such as
binary circularization and spin synchronization) in this regime would
be applicable to a number of other important systems, including
pre-main-sequence (hereafter PMS) binaries (Zahn \& Bouchet 1989).
The 51 Peg system has a very short-period (4.2 d) orbit, which makes
it particularly interesting in the context of tidal dissipation theory
(Rasio et al.~1996; Marcy et al.~1997).  

\subsection{Planet Formation}

The properties of the new planets are surprising (see Table~1).  Most
are Jupiter-mass objects in very tight circular orbits or in wider
eccentric orbits. The standard model for planet formation in our Solar
System cannot explain their orbital properties (see, e.g., Lissauer
1993; Boss 1995).  According to this standard model, planetary orbits
should be nearly circular, and giant planets should be found at large
distances ($\go1\,$AU) from the central star, where the temperature in
the protostellar nebula is low enough for icy materials to condense
(Boss 1995, 1996).  These simple predictions of the standard model for
the formation of the Solar System are at odds with the observed
parameters of most detected extrasolar planets.  With only one
exception (47 UMa), the new planets all come within $1\,$AU of the
central star. Three planets (51 Peg, $\tau$ Boo, $\upsilon$ And) are
in extremely tight circular orbits with periods of only a few days.
Two planets ($\rho^1$ Cnc and $\rho$ CrB) have circular orbits with
somewhat longer periods, of order tens of days.  Three companions with
wider orbits (\hbox{16 Cyg B,} 70 Vir, and HD 114762) have very large
eccentricities ($\sim 0.5$).  A number of different theoretical
scenarios have been proposed to explain the unexpected orbital
properties of these extrasolar planets.  Our calculations will provide
accurate stellar parameters that can be used to obtain better
constraints on these formation scenarios.

Four specific mechanisms have been suggested as possible processes to
bring a giant planet into a short-period orbit around its star.  One
mechanism is a secular interaction with a distant binary companion.
If the orbit of a wide binary is inclined relative to a planet's orbit
by more than $\sim 40^{\circ}$, the relative inclination of the binary
star can couple to a secular increase in the eccentricity in the
planet's orbit (Holman, Touma, \& Tremaine 1996; Mazeh, Krymolowski,
\& Rosenfeld 1996). The amplitude of the eccentricity perturbation
depends on the relative inclination of the orbits, but is independent
of the mass of the binary companion.  If the orbital planes have a
very high relative inclination, then an eccentricity approaching unity
can be induced.  In some cases this may cause a collision with the
star.  However, if dissipation is significant (in the star, in the
planet, or in a disk), then the orbit could circularize at a small
distance.  Similarly, if the primary star has a significant quadrupole
moment, then tidal dissipation in the star could stop the growth of
the eccentricity oscillations and drive a gradual decrease in the
semi-major axis of the planet's orbit (Kiseleva \& Eggleton 1997;
Eggleton \& Kiseleva 1997). This mechanism of secular perturbations
from a distant binary companion can also produce significant
eccentricities for the longer period extrasolar planets.  It is
interesting to note that four of the five planets with semi-major
axes, $a\lo 0.25$ AU are in wide binary systems (Hoffleit \& Warren
1991).  51 Peg has been searched extensively for a binary companion
(Marcy et al.~1997), but none has been found.  If the relative
inclination less than $\sim 40^\circ$, then the amplitude of the
eccentricity oscillations becomes sensitive to the mass of the
perturbing body, the ratio of the semi-major axes, and the
eccentricity of the outer orbit.  For a binary companion of stellar
mass, only a small range of parameters will produce large
eccentricities without disrupting the system.  The large mass ratio of
either star to the planet makes it easy for the system to be
disrupted.  Since the planet is the least massive of the three bodies
by many orders of magnitude, the planet is almost certainly the body
to be ejected, if the system dissociates.

Dissipation in the protostellar nebula is a second possible mechanism
for forming a short period planet. Since the orbital migration of the
planet would tend to accelerate with decreasing separation from the
star, the dissipation has to switch off at a critical
moment for a planet to end up so close to its parent star without
being disrupted.  Possible mechanisms for stopping the inward
migration include Roche lobe overflow and tidal coupling to a rapidly
rotating star (Trilling et al.~1998; Lin et al.~1996).  Another
possibility is that the migration stops when the planet arrives at the
inner edge of a disk limited by a magnetosphere around the star (Lin
et al.~1996).

A resonant interaction with a disk of planetesimals is another possible
source of orbital migration, but this requires a very large
protoplanetary disk mass if a $\sim 1 M_{\rm J}$ planet is to
migrate inwards all the way to $\sim0.1\,$AU (Murray et al.\
1998). The advantage, however, is that the migration is halted
naturally at short distances when the majority of perturbed
planetesimals collide with the star rather than escaping on nearly
parabolic orbits.  Wide eccentric orbits can also be produced for
planets more massive than $\sim3\,M_{\rm J}$.  It seems that such
a massive disk would be likely to produce more than one planet, and
thus it is important to understand how a second or third planet would
affect this scenario.

The fourth mechanism is based on dynamical instabilities in a system
originally containing multiple giant planets of comparable masses
(Rasio \& Ford 1996).  If either the orbital radii evolve secularly at
different rates (significant orbital migration is thought to have
occurred in the outer Solar System; see Goldreich \& Tremaine 1980;
Malhotra 1995) or if the masses increase significantly as the planets
accrete their gaseous envelopes (Lissauer 1993), then the orbits could
become unstable.  Alternatively, if either the second or third
mechanism for producing short period planets were truncated when the
disk mass dwindled, they could leave two planets in dynamically
unstable orbits.  Similarly, a wide binary companion could drive a
secular increase in the planets' eccentricity until they became
dynamically unstable.  In any case, the evolution can lead to a
dynamical instability of the orbits and a strong gravitational
interaction between two planets (Gladman 1993; Chambers, Wetherill, \&
Boss 1996). This interaction can lead to the ejection of one planet,
leaving the other in an eccentric orbit. If the pericenter distance of
the inner planet is sufficiently small, its orbit can later
circularize at an orbital separation of a few stellar radii (Rasio et
al.\ 1996).  This mechanism can produce eccentric systems in two
different ways.  After many encounters, one planet can be ejected from
the system leaving the other planet in an eccentric orbit by itself
(Rasio \& Ford 1996; Katz 1997).  In this scenario it should be
expected that the second planet has been ejected and is no longer in
the system (Black 1997).  Alternatively, if the two planets collide,
they can produce a more massive planet in an eccentric orbit (Rasio \&
Ford 1996; Lin \& Ida 1997).  Dynamical instabilities in systems
containing more than two giant planets of comparable masses have also
been studied.  Weidenschilling \& Marzari (1996) have obtained
numerical results for the case of 3 planets, while Lin \& Ida (1997)
performed simulations for systems containing up to 9 planets. In this
case successive mergers between two or more planets can lead to the
formation of a fairly massive ($\go10\,M_{\rm J}$) object in a
wide, eccentric orbit. While it is almost certain that this mechanism
operates in many systems with multiple planets, it is not clear that
it can produce a fraction of 51 Peg-like systems as large as observed.
Extensive numerical simulations will be necessary to obtain good
statistics on this theory (Ford \& Rasio 1998).  Observational
selection effects must also be better understood for a meaningful
comparison with the properties of the detected systems.

All these dynamical processes can also affect the evolution of the
central star.  Marcy et al.\ (1997) and Drake et al.\ (1998) argue
that the observed rapid rotation of $\tau$ Boo is likely caused by the
tidal interaction between the star and its close planetary companion
(our results do not support this interpretation; see \S 4.1).  Several
stars with short-period planets have high metallicities
($\left[\frac{Fe}{H}\right] \go 0.2$; see Gonzalez 1997a, 1998ab and
Table~2).  Gonzalez (1998a) proposes that their metallicities have been
enhanced by the accretion of high-Z material.  Alternatively, the
correlation could arise, if metal-rich stars have metal-rich disks
which are more likely to form planets.  Thus, understanding the
relationship between metallicity and the existence of short-period
planets may be important in constraining the mechanisms which produce
these planets.  We discuss this in more detail after constructing
models for these stars (See \S 4.2).

\section{Constructing Stellar Models}

\subsection{The Code}

We use the Yale Rotating Evolution Code in its non-rotating
mode to calculate stellar models. YREC is a Henyey code which solves
the equations of stellar structure in one dimension. The chemical
composition of each shell is updated separately using the nuclear
reaction rates of Bahcall \& Pinsonneault (1992). The initial chemical
mixture is the solar mixture of Anders \& Grevesse (1989), scaled to
match the metallicity of the star being modeled. For regions of the
star where $\log T (K) \geq 6$, all atoms are assumed to be fully
ionized. For regions where $\log T (K) \leq 5.5$, particle densities
are determined by solving the Saha equation for the single ionization
state of hydrogen and the metals, and for the single and double
ionization states of helium. In the transition region between these
two temperatures, both formulations are weighted with a ramp function
and averaged.  The equation of state includes both radiation pressure
and electron degeneracy pressure. We use the latest OPAL opacities
(Iglasias \& Rogers 1996) for the interior of the star down to
temperatures of $\log T (K) = 4$. For lower temperatures, we use the
low-temperature opacities of Alexander \& Ferguson (1994). We use an
Eddington T-$\tau$ relationship for the atmosphere, and, where
appropriate, the standard B\"{o}hm-Vitense mixing length theory of
convection (Cox 1968; B\"{o}hm-Vitense 1958) with the ratio of the
mixing length to the pressure scale height, $\alpha$=1.70.  This value
of $\alpha$, as well as the solar hydrogen abundance,
$X_{\odot}=0.697$, was obtained by calibrating models against
observations of the present day Sun using the aforementioned physics.

Astroseismology has made it possible to test models of stellar
interiors directly.  Models of the Sun (see, e.g., Guenther et
al.~1992) are constantly being refined with newer and more
sophisticated input physics, including diffusion of helium and heavy
elements (Guenther \& Demarque 1997) and detailed hydrodynamic
calculations of the surface convection zone (Demarque et al.~1997).
YREC has also been used to calculate models of other single field
stars.  Most recently, observations of p-mode oscillations in $\eta$
Bootis, a nearby subgiant, prompted detailed stellar modeling of that
star (Guenther \& Demarque 1996). This work forms the basis of our
method.  The modeled frequency spectra of the Sun and $\eta$ Bootis
agree quite well with observations.  YREC has also been used to model
Procyon (Guenther \& Demarque 1993) and both components of the binary
star $\alpha$ Centauri (Edmonds et al.~1992).  Although stellar models
have also predicted seismology spectra for Procyon and $\alpha$
Centauri, no convincing observations of the frequency spectra have yet
been made.

%\subsection{Uncertainty in Physics}

In constructing stellar models a variety of formalisms are available
for the treatment of the equation of state, the stellar atmosphere,
and convection.  We adopted formalisms which result in a good solar
model, and are standard enough to be applicable to all low-mass stars.
We have neglected non-standard effects such as diffusion of helium and
heavy elements, rotation, and magnetic fields since these processes do
not significantly change the overall properties of low-mass stars like
the sun. Solar models constructed by YREC with the set of standard
formalisms adopted here reproduce the observed solar p-mode spectrum
to within 1\% (Guenther et al.~1992).  The incorporation of newer
physics can reduce this difference by a few tenths of a percent.

We have chosen to use the Saha equation of state rather than 
MHD (Mihalas et al.~1988) or OPAL (Rogers \& Iglesias
1994). The MHD and OPAL equations of state are very similar to each
other, and differ from the YREC implementation of the Saha equation by
less than 1\%. The OPAL equation of state results in a slightly deeper
convection zone for the sun, and a worse fit to the observed p-mode
spectrum (Guenther et al.~1996) compared to the Saha equation of
state.

For the treatment of convective zones we have adopted the standard
B\"{o}hm-Vitense mixing length theory. Some other theories, such as
that of Canuto \& Mazzitelli (1992), produce solar models which have
p-mode spectra in slightly better agreement with observations to the
observed spectrum. Currently, a number of groups are using 3-D
numerical simulations of turbulent convection to develop a more
realistic description of stellar convective zones (e.g. Kim \& Chan
1998). However, the results of these simulations are not yet available
in sufficient detail to be incorporated into general stellar models.

The mixing length is chosen so that our solar model reproduces the
solar luminosity and solar radius at the solar age. Our best value of
the mixing length parameter (ratio of mixing length to the local
pressure scale height) is 1.7. To test the effect of the mixing length
on the stellar parameters, for several of our best models we varied
the mixing length from 1.4 to 2.0 (a large range). We found that the
luminosity of the star typically varied by $\sim$ 3\%, the radius by
$\sim$ 5\%, the mass of the convective envelope by $\sim$ 35\%, and
the radius of the convective envelope by $\sim$ 2\%.  We have also
performed a number of tests to check that the choice of atmosphere
model does not affect our results significantly. For examples, we find
that using the Kurucz tabulated atmospheres (Kurucz 1991) rather than
the Eddington T-$\tau$ relation increases the calculated effective
temperatures and increases the calculated radii by less than 1\%.  We
can test our complete set of physical parameters and assumptions by
comparing the calculated depth of the solar convection zone to that
derived from the observed solar p-mode spectrum. Christensen-Dalsgaard
et al. (1991) give the radius at the base of the solar convection zone
to be $0.713 \pm 0.003 $ R$_{\odot}$. Our calibrated model produces a
convection zone which begins at $0.647$ R$_{\odot}$, a difference of
$\sim 10\%$. In summary, we expect that the calculated stellar radii
and temperatures will be accurate to within about 5\%, and the
calculated size of the convection zones will be accurate to within
20\%, including all possible changes to the physics of the models.

\subsection{Iteration Method}

We constructed models of 0.6$M_{\odot}$, 0.8$M_{\odot}$,
1.0$M_{\odot}$, 1.2$M_{\odot}$, 1.4$M_{\odot}$, and 1.6$M_{\odot}$ PMS
stars by solving the Lane-Emden equation for a polytrope of index
$n=1.5$.  YREC evolved the models to the zero-age main sequence
(hereafter ZAMS).  These ZAMS models serve as the starting point for
all our subsequent modeling.

For each model we construct, we first specify a mass and metallicity
and then evolve a ZAMS model in an attempt to match a given effective
temperature $T_{\rm eff}$ and luminosity $L$.  First, YREC chooses a
ZAMS model with a similar mass and metallicity and scales the ZAMS
model to match the desired values of $M_{*}$ and $Z$ (indirectly
affecting $X$ and $Y$).  The scaled ZAMS model is numerically relaxed
before YREC begins to evolve the model to the desired radius
(determined from effective temperature and luminosity).  YREC iterates
the above procedure, returning to the ZAMS model, but rescaling to new
values $X$ and $Y$ so as to improve the match to the desired
luminosity and radius.

Thus, a given run fits a model to the desired luminosity and radius
(and hence temperature).  For such a run we hold $M_{*}$ and $Z$
constant, explicitly vary $Y$, and let $X$, the age, and the other
stellar parameters vary as a result of the changing composition.  This
procedure is repeated for several masses yielding a set of models
which share the same specified metallicity, luminosity, and
temperature, but differ in mass, age, composition, and other
parameters.

Since we vary the helium content in our models, it is possible to
obtain models that match the observed parameters but have unrealistic
compositions.  To select our final ``best''' model, we consider the
ratio \begin{equation} \frac{\delta Y}{\delta Z} \equiv \frac{ Y_{} -
Y_{\odot} }{ Z_{} - Z_{\odot} }, \end{equation} where $Y_{}$ and
$Z_{}$ are the helium and heavy element abundances of our model.
Based on studies of galactic evolution, including H II regions (Pagel
et al.~1992) and low-metallicity blue compact galaxies (Izotov et
al.~1997), we impose the canonical constraint $\frac{\delta Y}{\delta
Z} = 2.5 \pm 1$ (Bressan et al.~1994; Edvardsson et al.~1993).  For
our ``best'' model, we interpolate to find a mass and age
corresponding to $\frac{\delta Y}{\delta Z}= 2.5$ for the specified
parameters.  Similarly, we present models for $\frac{\delta Y}{\delta
Z}$ equal to $1.5$ and $3.5$ from which we determine theoretical error
bars given a set of assumed observational parameters.  Due to the
physical scatter and the possibility of systematic errors in the
determination of $\frac{\delta Y}{\delta Z}$, we also include models
for $\frac{\delta Y}{\delta Z}$ equal to $5.0$ and $0.0$.  The latter
may also be useful for comparisons with studies in which $Y=Y_{\odot}$
is assumed.  Fortunately, the theoretical uncertainties derived from
the variation of $\frac{\delta Y}{\delta Z}$ are, in general, smaller
than the uncertainties in the stellar observations, as will be
discussed in the next section.  The above procedure is repeated for
each set of observed parameters, $Z$, $T_{\rm eff}$, and $L$.  For
some sets of observed parameters, YREC is unable to converge on a
single self-consistent model.

\subsection{Observational Data}

The above procedure for constructing a set of models to match a real
star requires three basic input parameters from observations.  The
luminosity, $L$, effective temperature, $T_{\rm eff}$, and metallicity,
$Z$, must be known accurately.  The observational uncertainty in these
quantities limits the accuracy with which we can derive the values of
other stellar parameters.  As more observational data become
available, these can be used to constrain our models more tightly.

\subsubsection{Luminosities}

To minimize systematic errors, we consistently use the stellar
luminosities and parallaxes obtained from the Hipparcos catalog.  The
recent release of the Hipparcos data (ESA 1997, Perryman et al.~1997)
is extremely useful for this type of study, since accurate distances
to all the relevant stars are now available.  The Hipparcos parallax
data have greatly improved astronomical distances and tightly
constrained luminosities, nearly removing an entire degree of freedom
from the models.  Since only the brightest (and therefore the closest)
solar-type stars are targets for present spectroscopic surveys, all
the solar-type stars presently known to have planets are nearby and
have distances directly determined by Hipparcos parallax data. This
distance is combined with the Hipparcos apparent visual magnitude to
obtain a visual luminosity.

One minor complication is the application of bolometric corrections to
convert visual luminosities to integrated luminosities. We estimate
bolometric corrections by interpolating in [Fe/H], $\log g$, and
$T_{\rm eff}$ across color calibration grids used for the Yale
Isochrones (Green et al.~1987).  This method of estimating bolometric
corrections does introduce a dependence of our results on $\log g$.
However, bolometric corrections depend only weakly on $\log g$, and
the uncertainty in $\log g$ hardly introduces any uncertainty in the
bolometric correction.  Comparing our bolometric corrections with other
tables indicates that there is a $\sim 5\%$ systematic uncertainty in
our luminosities due to the potential systematic error in the
bolometric corrections.  From our full set of models we see that the
uncertainty in the integrated luminosity is normally insignificant
compared to other sources of error.  

\subsubsection{Spectroscopic Data}

In addition to the luminosity, our models also requires a knowledge of each
star's metallicity and effective temperature.  Effective temperatures
and metallicities can be obtained from either photometric or
spectroscopic data, although the latter are normally more accurate.
In this paper we have chosen values for the necessary parameters from
the literature.  To minimize systematic errors, we decided to use 
temperatures, metallicities, and surface gravities obtained from a
single source, the recent spectroscopic observations of all the
solar-like stars with planets by Gonzalez (1997ab, 1998ab).  His 
determinations of $T_{\rm eff}$, $\log g$, [Fe/H], and the
microturbulence parameter $\zeta_t$, are from self-consistent
iterative solutions which matched high resolution spectra of Fe I and
Fe II lines to Kurucz (1993) model atmospheres.  Gonzalez (1998) quotes
typical errors of 75 K, 0.06, and 0.05 for $T_{\rm eff}$, [Fe/H],
and $\log g$, respectively.  These formal errors may be somewhat
smaller than the actual uncertainty in the values, but the general
agreement with other observations is reassuring.  We construct
additional models when other observations significantly deviate from
the Gonzalez results, or when we cannot construct models simultaneously
matching all the observed parameters.  We summarize the observed
parameters of stars with planets in Table 2.  We will list other
determinations of the stellar parameters, $T_{\rm eff}$, [Fe/H],
$\log g$, and $v \sin i$, as we discuss each star individually.

\subsubsection{Other Observational Parameters}

Several other stellar parameters can be measured observationally.  For
example, $\log g$ is routinely measured spectroscopically.  While
determinations of $\log g$ serve as a useful tool for purposes such as
determining bolometric corrections (see \S 2.3.1), present
determinations ($\sigma_{\log g (cgs)} \sim 0.1$) provide only loose
constraints ($\sim 25\%$) on $M_{*}/R_{*}^2$.  Thus, we use
measurements of $\log g$ merely as a consistency check for our models.

Analysis of Ca II H and K emission can be used to detect the stellar
rotation period.  When a rotation period is not detected, it can be
predicted using an empirical relation between the Ca II flux and the
rotation period (Noyes et al.~1984).  The rotation period can be
combined with the radius to yield an equatorial velocity, $v_{\rm eq}$.
This can be compared with the observed $v \sin i$ to
yield the inclination angle between the star's equator and the line of
sight.  If the angle between the star's equator and the planet's
orbital plane is presumed small, one can determine the planet's actual
mass, $m$, from its minimum mass, $m \sin i$.  Unfortunately, $v \sin
i$ is very difficult to determine observationally, as it is delicately
coupled to the macroscopic turbulence parameter.  We list
only recent determinations of $v \sin i$ and take even these somewhat
cautiously.  Empirically, both the level of Ca II H and K activity and
the stellar rotation frequency have been found to decrease with age and
this relationship can be used to estimate the age of the star
(Baliunas et al.~1995; Soderblom et al.~1991).  We will compare our
determinations of the age with those predicted by the activity-age
relation.

Photometric determinations of the angular diameter are possible with
the Barnes and Evans relationship (Barnes, Evans, \& Moffet 1978;
Moffett \& Barnes 1979) or the more recent infrared flux method
(Blackwell et al.~1990).  Combined with parallax measurements, these
yield a stellar radius.  Given the relatively high uncertainties
associated with these methods, we also use these observations only as
a consistency check.

\section{Results}

For each star we have computed a grid of models surrounding the
observed values of $L$, $T_{\rm eff}$, and [Fe/H].  As more
observations become available, it is hoped that our large grids of
models will allow observers to translate their values of the observed
parameters into physical parameters such as mass, radius, age, and
size of the convective zone.  In general, interpolating across grids
of stellar models can be inaccurate as non-linearities become
significant.  However, we have calculated a fine enough grid of models
for each star such that interpolation should yield reasonable results.
The set of models printed here is only a small subset of our large
grid of models.  Here we present models corresponding the adopted
values of $T_{\rm eff}$, [Fe/H], and $L$ and models corresponding to a
$1-\sigma$ uncertainty in any one of these parameters.  For each model
we include the mass $M_{*}$, the age, the mass of the convective
envelope $M_{\rm ce}$, the radius of the convective envelope $R_{\rm
ce}$, the pressure scale height at the base of the convective envelope
PSH, the radius $R_{*}$, and the surface gravity $\log g$.  The full
set of models will be made available electronically.

% *******************************************************************
% ********************  Note to the Editor  *************************
% *******************************************************************
%    How should we go about doing this?  An ftp site, a CD, CDS?
% *******************************************************************

We summarize our results in Table 3 with determinations of the mass,
radius, age, size of convective envelope, and the eddy turnover
timescale at the base of the convective envelope for each star (See \S
4.1).  The error bars indicate the range of values obtained in models
with $\frac{\delta Y}{\delta Z} \in \left[ 1.5, 3.5 \right]$ and each
of the input parameters, $L_{*}$, $T_{\rm eff}$, and [Fe/H] varied
within the ranges specified below for each star.  These error bars do
not include the uncertainty in the choice of physical models discussed
in \S 2.1.  For our best model of each star we also list the
dimensionless gyration radius of the star, the dimensionless gyration
radius of the convective envelope (See \S 4.1) and the mass of the
convective envelope at the ZAMS.

\subsection{51~Pegasi}

The planet around 51 Pegasi was discovered by Mayor \& Queloz (1995)
and has since been confirmed by both Marcy et al.~(1997) and Horner et
al.~(1997).  The SFSU team has the cleanest data, giving an rms
scatter of 5.2 m s$^{-1}$ about a Keplerian fit with a semi-amplitude
of $55.9\pm0.8$ m s$^{-1}$.  They calculate an orbital period of
$4.2311\pm0.0005$ d and an eccentricity of $0.012\pm0.010$.  The
possibility of a second companion has been carefully examined.  The
SFSU planet search should already be able to detect planets with $m
\sin i \simeq 1 M_{J}$ within 2 AU.  A longer temporal baseline of
observations is necessary to extend this limit (Marcy et al.~1997).
Several teams have searched for periodic spectral line bisector
variations.  Although there is no convincing evidence (Horner et
al.~1997; Hatzes, Cochran, \& Cohns-Krull 1997; Gray 1998; Hatzes,
Cochran, \& Bakker 1998), the possibility of exciting non-radial
stellar pulsations in 51 Peg-like systems remains interesting (Willems
et al.  1997; Terquem et al.~1998), and these may become detectable in
the future.

We calculated our models with the values from Gonzalez (1998a): $T_{\rm
eff}=5750\pm75$ K and [Fe/H]$=0.21\pm0.06$.  The consistent
determinations of these parameters from many observations (See Table
3) is comforting.  We obtain a best model of 51 Peg with a mass of
$1.05^{+0.09}_{-0.08} M_{\odot}$, radius of
$1.16\pm 0.07 R_{\odot}$, and an age of $7.6^{+4.0}_{-5.1}$
Gyr.  We find that 51 Peg has a convective envelope ($M_{ce}
\simeq 0.023^{+0.007}_{-0.006} M_{\odot}$) only slightly larger than
that of the sun ($M_{ce} \simeq 0.0174 M_{\odot}$).  The eddy turnover
timescale calculated as in Rasio et al.~(1996) is $\tau_{ce} \simeq
18.6\pm2.5$ d which is close to the expected value for
main-sequence solar-type stars, $\tau_{ce} \sim 20$ d.  In addition,
we can follow the history of the convective envelope.  We find that
the mass of convective envelope on the ZAMS was $M_{\rm ce,ZAMS}\simeq
0.037 M_{\odot}$.

As a consistency check, we compare our predicted value of $\log g$
(cgs) = $4.33\pm0.09$ with those determined observationally.  We find
general agreement, especially when we consider that several of the
lower observed values have already been criticized in the literature.
Fuhrmann, Pfeiffer, \& Bernkopf (1997) suggest that the low value of
the Gratton, Carrenton, \& Castelli (1996) study is a consequence of
their low $T_{\rm eff}$, and that the McWilliam (1990) result
should be discounted, as it depended on an earlier misclassification
of 51 Peg as a subgiant.  Indeed, the reclassification of 51 Peg as a
dwarf has been verified by multiple observations.  Fuhrmann et
al.~(1997) also discount the Xu (1991) value, as it is based on low
resolution spectra.  Finally, the Edvardsson et al.~(1993a,b) result is
superseded by their own more recent observations which appear in
Tomkin et al.~(1997).  The remaining determinations are all consistent
with our models.

The activity-age relation predicts an age of $10$ Gyr consistent with
our result for 51 Peg (Baliunas et al.~1997), and a rotation period of
29.6 d, while a rotation period near 37 d has been observed (Baliunas
et al.~1996). Combining a rotation period of $34\pm4$ d with our
radius, we compute an equatorial velocity, $v_{\rm eq}=1.78\pm0.23$ km
s$^{-1}$.  This can be combined with observed values of $v \sin i$ to
provide estimates of the inclination and hence the actual companion
mass.  The determination of $v \sin i = 1.4\pm0.3$ km s$^{-1}$ by
Gonzalez (1998) is quite consistent with our calculated $v_{\rm eq}$,
implying $\sin i=0.8\pm0.2$ and $m \simeq 0.59\pm0.15 M_{\rm J}$.
However, most recent determinations of $v \sin i$ are larger than the
above computed $v_{\rm eq}$ (See Table 4). It should be noted that
both the values of Hatzes et al.~(1997) and Francois et al.~(1996)
have already been corrected according to Gonzalez (1998).  These
measurements are either inconsistent or barely consistent with the
radius and period, suggesting that $\sin i \sim 1$ and thus the
companion mass is not very different from the minimum companion mass.

\subsection{$\tau$ Bootis}

The SFSU planet search discovered a planet around $\tau$ Boo in a
$3.3128\pm0.0002$ d near-circular orbit.  There is an rms scatter of
13.9 m s$^{-1}$ with occasional episodes of greater scatter about the
Keplerian fit with semi-amplitude $469\pm5$ m s$^{-1}$.  The scatter is
significantly above the instrumental error, but cannot be explained by
the presence of a second planet (Butler \& Marcy 1996). Several teams
have searched $\tau$ Boo for periodic spectral line bisector
variations, but none have been found (See \S 3.1).

We started with models based on the values of $T_{\rm eff}$ and [Fe/H]
from the recent spectroscopic work of Gonzalez (1997a).  In the
process, we found that our models favored lower values of $T_{\rm
eff}$ and [Fe/H], and so we considered other determinations of $T_{\rm
eff}$ and [Fe/H]. Upon reanalysis of his own data, Gonzalez has
refined his estimate of $T_{\rm eff}$ to $6550\pm100$ K (Gonzalez
1997b).  He claims that his data might be consistent with $T_{\rm
eff}=6400$ K, but that they are inconsistent with $T_{\rm eff}=6300$
K.  There may be a significant difference between spectroscopic and
photometric determinations of $T_{\rm eff}$ and [Fe/H] (Gonzalez
1997a).  The small number of metal-rich stars used in the calibration
of photometric estimators is one possible explanation for the
difference in spectroscopic and photometric estimates for $\tau$ Boo
(Gonzalez 1997a).  Alternatively, if $\tau$ Boo has a convective
envelope, the tidal torque from a planet in a 3.3 d orbit could drive
the envelope to rotate much more rapidly than normal, thereby driving
activity on the stellar surface and altering the observed spectral
characteristics of the star.  However there is a more likely
explanation: the high X-ray luminosity, radial velocity noise, young
age, and observed rotation period are all consistent will $\tau$ Boo
being a young star that is still rotating rapidly.  A rotation period
as short as 3.3 d can cause significant discrepancies between
spectroscopic and photometric determinations of $T_{\rm eff}$ and
[Fe/H]. The rapid rotation will broaden spectral lines and slightly
redden the color. This, like the high metallicity of $\tau$ Boo, can
affect the continuum and thus all the line depths and
widths. Photometric estimates for such rapidly rotating stars can also
be adversely affected.

We explored a range of $T_{\rm eff}$ and [Fe/H] significantly larger
than the uncertainties in the Gonzalez (1997a,b) observations require.
YREC was unable to construct any self-consistent models of $\tau$ Boo
in the range of 1.0 to 1.8 $M_{\odot}$ using $T_{\rm eff} \ge 6600$ K
and [Fe/H]$\go 0.13$, the spectroscopic value from Gonzalez (1997a).
If we adopt $T_{\rm eff}=6400\pm100$ and [Fe/H]$=0.25\pm0.09$, then we
arrive at a mass of $1.37\pm 0.08 M_{\odot}$, a radius of
$1.41^{+0.10}_{-0.09} R_{\odot}$ and an age of $1.2^{+1.2}_{-0.8}$
Gyr.  Our models predict $\log g=4.27^{+0.05}_{-0.07}$, which is in agreement
with most of the observations, but slightly inconsistent with the
Gonzalez (1997a) value of $4.5\pm 0.15$.  The most interesting result
from our models of $\tau$ Boo is the likely absence of a convective
envelope.  We find thin convective envelopes only in models with
$T_{\rm eff} \lo 6350$ K or [Fe/H]$\lo 0.07$ (See Table 7).

The activity-age relation predicts an age of $\sim 2$ Gyr (Baliunas et
al.~1997), consistent with our age.  The Mount Wilson HK Project has
revealed that $\tau$ Boo has chromospheric emission periods of
approximately $3.5\pm0.5$ d, 117 d, and 11.6 yr (Baliunas et
al.~1997).  The shortest period is believed to be the rotation period,
which corresponds very nearly to the orbital period of the planet.
That has led Marcy et al.~(1997) and Drake et al.~(1998) to suggest
the possibility that the star may have been tidally spun up.  If we
assume the star's rotation period is synchronized with the planet's
orbital period and combine the 3.3 d rotation period with our
determinations of radius, then we calculate $v_{\rm eq} =
21.7^{+1.1}_{-1.4}$ km s$^{-1}$.  This is consistent with all
determinations of $v \sin i$ and suggests that $\sin i\sim
0.67^{+0.07}_{-0.06}$, implying $m=7.1\pm0.8 M_{\rm J}$.

\subsection{$\upsilon$ Andromedae}

The SFSU team detected a companion to $\upsilon$ And in a $4.611\pm
0.005$ d near-circular orbit.  There is a 12.1 m s$^{-1}$ rms scatter
about the Keplerian fit of semi-amplitude $74.1\pm0.4$ m s$^{-1}$.  This
scatter is well above the instrumental error and could be either 
intrinsic to the star or indicative of a second planet.
Early observations reported a significant eccentricity of
$0.109\pm0.040$ (Butler \& Marcy 1996), but the long-term trend in the
residuals complicates the measurement of the eccentricity.  Thus, the
current uncertainty is likely larger than the quoted error bar (Marcy
1998).

Based on the observations of Gonzalez (1997a), we calculate a mass of
$1.34^{+0.07}_{-0.12} M_{\odot}$, radius of
$1.56^{+0.11}_{-0.10} R_{\odot}$, and age of $2.6^{+2.1}_{-1.0}$
Gyr.  The convective envelope is very shallow with $M_{ce} \simeq
0.002^{+0.003}_{-0.002} M_{\odot}$ and the eddy turnover time is
correspondingly short, $\tau_{ce} \simeq 6.8^{+2.3}_{-6.8}$ d.

The activity-age relation predicts an age of $5$ Gyr, consistent with
our results (Baliunas et al.~1997).  Based on Ca II emission,
$\upsilon$ And is expected to have a rotation period of $\sim 12$ d
(Baliunas et al.~1997).  If accurate, one can infer an equatorial
velocity $v_{\rm eq} \sim 6.6$ km s$^{-1}$.  This value is low
compared to observed values of $v \sin i \sim 9$ km s$^{-1}$ (See
Table 8).  This suggests a significant error in either the observed
value of $v \sin i$ or the estimated rotation period.  If we were to
take the reported $v \sin i$ at face value, then we would expect a
rotation period $\lo 6$ d.

Our models give $\log g=4.18^{+.07}_{-.10}$, which is consistent with
all but one of the observed values of $\log g$.  Blackwell et
al.~(1990) have used the infrared flux method to calculate an angular
diameter of $1.103\pm0.044$ mas, which, combined with the Hipparcos
parallax, gives $R_{*}=1.60\pm0.08 R_{\odot}$, in agreement with our
models.  This corrects some previous estimates based on the Blackwell
et al.~(1990) data and older parallax data (most notably $56.8\pm4.1$
mas from van Altena et al.~1995 which led to incorrect estimates of
$T_{\rm eff}$ and the radius).

\subsection{$\rho^1$ Cancri}

The SFSU team has discovered a companion around $\rho^1$ Cnc in a
near-circular orbit of period $14.648\pm0.0009$ d.  They find a
Keplerian fit to the radial velocity with a semi-amplitude of
$77.1\pm0.9$ m s$^{-1}$ and a rms scatter of $12.0$ m s$^{-1}$.  In
addition, there appears to be a long-term trend in the residuals,
possibly indicative of a second planet with mass $\sim 10 M_{\rm
J}$ and a period of $\sim 20$ yr (Marcy \& Butler 1998). Previously,
McAlister et al.~(1993) had searched for a luminous companion in a
similar orbit using speckle observations, but found none.

Initially, we attempted to construct models based on the stellar
parameters measured by Gonzalez (1998ab).  However, we found that this
value of the temperature, $T_{\rm eff}=5150\pm75$ K, is too low to
match any model of $\rho^1$ Cnc, even when evolved for 20 Gyr.

There are several possible explanations for the apparent inconsistency
of the various available data for $\rho^1$ Cnc.  One possibility is
that $\rho^1$ Cnc is actually a subgiant.  Indeed, $\rho^1$ Cnc's spectra closely
matches that of $\delta$ Eri, a subgiant with [Fe/H]$\simeq -0.15$
(Baliunas et al.~1997)\footnote{Cayrel de Strobel et al. 1997 lists
many values of [Fe/H] for $\delta$ Eri, ranging from -0.27 to 0.33.}.
Gonzalez (1998a) agrees that the spectrum is suggestive of a subgiant,
as is the low surface gravity.  However our models, as well as the
isochrones used by Gonzalez (1998a), would predict an extremely large
age, of $\go 12$ Gyr.

Gonzalez (1998) proposes another possible explanation: the
$\rho^1$ Cnc system may be an unresolved stellar binary viewed nearly
face-on, despite the {\it a priori} low probability of such a viewing
angle.  Gonzalez (1998) suggests monitoring the line profile
variations to test this hypothesis.  We discuss this possibility
further in \S 4.1.2, and find this hypothesis unattractive.

Another possible explanation is an error in the stellar models.  The
observed value of [Fe/H], $+0.45\pm0.03$ for $\rho^1$ Cnc is quite
extreme and the use of $\delta Y/\delta Z=2.5
\pm 1.0$ to determine Y for our models (as well as the isochrones of
Schaller et al.~(1992) and Scaerer et al.~(1993a,b) used by Gonzalez
1998) may not be appropriate for such high metallicity stars.  While this
constraint on $\delta Y/\delta Z$ is believed to be reasonable for
most solar-like stars, the calibration is based on lower metallicities
and this linear relationship may not be adequate for high values of Z.

Finally, it is possible that the observations have larger errors than
those quoted.  In particular, a higher $T_{\rm eff}$ would restore
consistency.  However, since the observed value of [Fe/H] is
correlated with $T_{\rm eff}$, the observed metallicity of $\rho^1$
Cnc would then increase from its (already high) quoted value (Gonzalez
1996, 1998ab).  If we adopt $T_{\rm eff}=5300\pm75$ and
[Fe/H]$=0.45\pm0.03$, then we find a mass of $0.95^{+0.11}_{-0.09}
M_{\odot}$ and a radius of $0.93^{+0.02}_{-0.03} R_{\odot}$.  The
convective envelope ($M_{ce} \simeq 0.046^{+0.004}_{-0.006}
M_{\odot}$) is significantly larger than solar and the eddy turnover
time is slightly longer, $\tau_c \simeq 26.7^{+1.2}_{-2.2}$ d.  The
age is very sensitive to $Y$ and thus is not well constrained.  This
explanation of our difficulties in modeling $\rho^1$ Cnc is supported
by the more recent observations of $\rho^1$ Cnc by Fuhrmann et
al.~(1998) and Gonzalez (1998b).  Given these recent observations and
the difficulties with the alternative explanations, we find this
explanation the most attractive.  Although the Fuhrmann et al.~(1998)
and Gonzalez (1998b) observations consistently support a larger
$T_{\rm eff}$, the unknown helium fraction still prevent us from
obtaining an accurate determination of the age.  Fuhrmann et
al.~(1998) obtained a maximum age based on their own observations.  In
addition to their observations having the largest $T_{\rm eff}$, they
have assumed a solar helium abundance.  If we use their observational
data and assume a solar helium abundance, then we obtain a similar
upper limit on the age.  However, if we use their observational data,
but assume $\delta Y/\delta Z=2.5 \pm 1.0$, then we can no longer
impose such a constraint.  Thus, their determination of a maximum age
for $\rho^1$ Cnc is only valid if $\rho^1$ Cnc has a near solar helium
abundance.

In light of the problems modeling $\rho^1$ Cnc, we consider several
other pieces of observational data.  Our models give $\log
g=4.50^{+0.04}_{-0.07}$, significantly higher than the Gonzalez (1996,
1998) value of $4.15\pm0.05$, but in agreement with other observations
(See Table 10).  This is expected since $\log g$ is correlated with
$T_{\rm eff}$. The activity-age relation suggests an age of $\sim5$
Gyr (Baliunas et al.~1997).  The observed rotation period of 41.7 d
(Baliunas et al.~1997) can be combined with our radius to obtain
$v_{\rm eq} = 1.12\pm0.03$ km s$^{-1}$, barely consistent
with the observed value of $v \sin i$, $1.4\pm 0.5$ km s$^{-1}$.  The
Barnes and Evans relationship (Barnes et al.~1978) gives an angular
diameter of $0.790\pm0.032$ mas, which we combine with the Hipparcos
parallax to calculate a radius of $1.06\pm0.04 R_{\odot}$, 
larger than the radii we find in our models.

\subsection{$\rho$ Coronae Borealis}

The Advanced Fiber Optic Echelle (AFOE) spectrograph team discovered a
companion orbiting $\rho$ CrB in a $39.645\pm0.088$ d near-circular
orbit.  They fit a Keplerian curve of semi-amplitude $67.4\pm2.2$
m s$^{-1}$ to their data leaving a rms scatter of 9.2 m s$^{-1}$ (Noyes et
al.~1997).  More recent data indicate that the orbit may have an
eccentricity of $0.15\pm0.03$ (Marcy 1998).

We find the metallicity [Fe/H] $=-0.29\pm0.05$ quoted by Gonzalez
(1998a) to be only marginally consistent with our models.  We find
consistent models only at the upper end of the quoted error bar.
Using the observations of Gonzalez (1998a), but with a slightly higher
[Fe/H]$=-0.23\pm0.06$, we calculate a mass of $0.89^{+0.05}_{-0.04}
M_{\odot}$, radius of $1.35^{+0.09}_{-0.08} R_{\odot}$, and age of
$14.1^{+2.0}_{-2.4}$ Gyr.  We calculate $M_{ce}\simeq
0.033^{+0.011}_{-0.009}$ and $\tau_c \simeq 21.5^{+2.9}_{-2.8}$. We
find general agreement between our computed $\log g=4.13^{+0.07}_{-0.06}$ and
the observed values (See Table 12).

The activity-age relation predicts an age of $6$ Gyr (Noyes et
al.~1997), significantly younger than in our models. While our old age is
worrisome, it is not extremely sensitive to any of the input parameters.
For example, insisting that the age is $\lo 10$Gyr, would require a
$2-\sigma$ error in both the temperature and the metallicity.  No
rotation has been observed, but based on the Ca II flux, the rotation
period is predicted to be $\sim 20$ d (Noyes et al.~1997).  Combining
this period and our radius, we compute $v_{\rm eq} \sim 3.4$ km
s$^{-1}$.  This is consistent with the observed values of $v \sin i
\sim 1.5$ km s$^{-1}$, but an accurate estimate of $\sin i$
cannot be obtained given the large uncertainty in the rotation period.

\section{Implications for Planet Formation and Evolution}

\subsection{Tidal Dissipation}

%\subsubsection{Orbital Decay and Circularization}

Our models do not change the main conclusions of Rasio et al.~(1996)
concerning the importance of orbital decay driven by tidal dissipation
in the 51 Peg system, but they do alter some of the quantitative
estimates (See Fig.~1). The planets around $\tau$ Boo and $\upsilon$
And have longer timescales for orbital decay, since, as our models
show, these stars have at most a very shallow convective envelope.
The larger semi-major axes of the planets around $\rho^1$ Cnc and
$\rho$ CrB increase their timescale for orbital decay.  Thus, our
models show that all the known extrasolar planets have stable orbits,
in the sense that the orbital decay timescale is long compared to the
main-sequence lifetime of their star.

Since orbital circularization is dominated by tidal dissipation {\em
in the planet}, the only relevant stellar parameter for
circularization is the stellar mass, as it affects our determination
of the mass of the planet and the semi-major axis from observations.
Parameters for the planet, such as the radius and the tidal
dissipation factor, $Q$, are less certain, but we can estimate them
using Jupiter as a guide (Rasio et al.~1996; Lubow, Livio, \& Tout 1997). Tidal
dissipation in the planet could circularize an eccentric orbit for the
51 Peg, $\tau$ Boo, and $\upsilon$ And systems, and perhaps the
$\rho^1$ Cnc system as well.  Tidal dissipation in the planet could
not have circularized an eccentric orbit in the case of $\rho$ CrB.

%\subsubsection{Maximum Companion Masses}

Tidal dissipation in the star also tends to spin up the star towards
synchronization with the orbital motion of the planet.  since radial
velocity surveys can only determine the $v \sin i$, some of the
detected systems may actually contain low-mass stellar companions in
orbits that happen to be viewed nearly face-on ($\sin i \ll 1$).
However, the probability of finding a $1 M_{\odot}$ companion with $m
\sin i$ as low as 1 $M_{\rm J}$ or 10 $M_{\rm J}$ is only one in $\sim
2 \cdot 10^6$ or $\sim 2 \cdot 10^4$, respectively.  Since the SFSU
planet search has already found 6 systems by monitoring a target list
of only 120 stars, the spectroscopic binary explanation is extremely
unlikely, but cannot be strictly ruled out.  In principle, astrometry
could be used to constrain the maximum companion mass.  However, this
is extremely difficult for short-period systems, since the position
wobble is so small.  For example, assuming a low mass companion to 51
Peg, the amplitude of the induced wobble is only $3 \cdot 10^{-3} /
\sin i$ mas.  Sophisticated space-based interferometers will be
necessary to measure such small effects.  Absence of X-ray activity
can also be used to confirm that a companion is substellar, since
close spectroscopic binaries are usually observed to have significant
X-ray luminosities.  However, a young stellar age or rapid rotation
also correlate with strength of X-ray emission (see e.g., Pravdo et
al.~1996).

Here we focus on the tidal constraints, which can place an
upper limit on the mass of companions in short-period systems.  For
the planet to survive, the orbital decay timescale must be large
compared to the age of the system.  If the companion has not yet spun
up the stellar rotation period to match its orbital period, then
another constraint can be imposed.  The maximum companion mass, as
determined from the timescales for either orbital decay or spin-up,
can be calculated from the stellar parameters.  Following standard
tidal dissipation theory (e.g., Zahn 1977; Zahn \& Bouchet 1989; Rasio
et al.~1996), we calculate the timescale for a planet to spin up the
whole star\footnote{ Note the obvious typographical error in Eq.~2 of
Rasio et al.~(1996).  Here we use standard tidal dissipation theory
based on the weak friction approximation with eddy viscosity as the
dominant dissipation mechanism in stars with convective envelopes
(Zahn 1977; Rasio et al.~1996).  Although the theory remains
controversial (see, e.g., Goodman \& Dickson 1998; Tassoul \& Tassoul
1997), there is empirical support for some of its main predictions
(Verbunt \& Phinney 1995; Zahn 1992).} as
\begin{equation}
\tau_{\rm su}^{*} = \frac{k_{*}^2 M_{*} R_{*}^2}{ma^2} \tau_{\rm a}, 
\end{equation}
where $k_{*}=\left( I_{*} / M_{*} R_{*}^2 \right) ^{1/2}$ is the dimensionless
gyration radius, $M_{*}$ is the mass of the star, $R_{*}$ is the stellar
radius, $m$ is the mass of the planet, $a$ is the semi-major
axis, and $\tau_{\rm a}$ is the timescale for orbital decay,
$$ \tau_{\rm a}^{-1} = \frac{f}{\tau_c} \frac{M_{\rm ce}}{M_{*}} q \left( 1+q \right) \left( \frac{R_*}{a} \right)^8.$$
Here $q=m/M_{*}$ is the mass ratio, $f$ is a numerical factor of order
unity, and $\tau_{\rm c}$ is the eddy turnover timescale.  Following Rasio
et al.~1996, we estimate $\tau_{\rm c}$ by
\begin{equation}
\tau_{\rm c} = \left[ \frac{ M_{\rm ce} R_{\rm ce} \left( R_{*}- R_{\rm ce} \right) }{3L_{*}} \right] ^{1/3},
\end{equation}
where $L_{*}$ is the stellar luminosity, $M_{\rm ce}$ is the mass
of the convective envelope, and $R_{\rm ce}$ is the radius at the
base of the convective envelope.  When the tidal pumping period (half
the orbital period) is small compared to the eddy turnover time in the
convective envelope, we expect that the efficiency of the tidal
dissipation will be reduced.  Thus, for fast tidal pumping, we reduce
the factor $f$ according to
\begin{equation}
f = f' {\rm min} \left[ 1, \left(\frac{P}{2\tau_{\rm c}} \right)^2 \right].
\end{equation}
Since the exact form of this correction is rather uncertain (see
Goodman \& Oh 1997 and references therein), we compute tidal decay and
spin-up timescales using both $f=1$ and $f$ as given by Eq.~4 with $f'=1$.

An even stronger constraint can be imposed if the spin up of the
convective envelope is assumed to occur independently from the
interior.  In the limit of a thin convective shell, the spin-up
timescale becomes
\begin{equation}
\tau_{\rm su}^{\rm ce} \simeq \frac{ M_{ce} R_{*}^2}{ma^2} \tau_{\rm a} \simeq \frac{k_{\rm ce}^{2}}{k_{*}^2} \tau_{\rm su}^{*}.
\end{equation}
In reality, the true
spin-up timescale is likely to be somewhere between $\tau_{\rm su}^{\rm ce}$ and $\tau_{\rm su}^{*}$.

We now consider the importance of tidal effects in each system individually.  

% Begin 51 Peg

{\bf 51 Peg}. It has been suggested that the 51 Peg system could
contain a stellar-mass companion, but that the system is being viewed
nearly face-on (Kubat, Holmgren, \& Rentzsch-Holm 1998; Imbert \&
Prevot 1998).  Astrometric measurements with Hipparcos impose an upper
limit of $\sim 500 M_{\rm J} \sim 0.5 M_{\odot}$ (Perryman et
al.~1997).  Diffraction limited ($\sim 0.05$ arcsec) infrared speckle
imaging with the Keck telescope searched mainly for M dwarf companions
and put limits on any such companions.  These detection limits give a
range of maximum orbital separations from 0.75 AU for M3 dwarfs to 23
AU for M7 dwarfs (Marcy et al.~1997).  The Palomar Testbed
Interferometer has also searched the 51 Peg system and determined a
maximum companion mass of $0.22 M_{\odot}$ (Boden et al.~1998). Low
levels of X-ray emission also indicate that the 51 Peg system is
unlikely have a stellar-mass companion (Pravdo et al.~1996).

We can put another constraint on the maximum companion mass based on
the slow rotation rate of the star implying $\tau_{\rm su} \gg
\tau_{\rm MS}$.  We calculate $\tau^{*}_{\rm su} \sin^2 i
\simeq 8.4 \cdot 10^{13}$ yr.  Comparing this to the best model age of
51 Peg, we get a maximum companion mass of $46 M_{J}$.  If we set
$f=1$, ignoring the reduction in efficiency from the rapid pumping
period, then we obtain $\tau^{*'}_{\rm su} \sin^2 i \simeq 1.1
\cdot 10^{12}$ yr, which yields a maximum mass of $5.5 M_{\rm J}$.
We also calculate $\tau^{\rm ce}_{\rm su} \sin^2 i \simeq 1.2
\cdot 10^{13}$ yr and $\tau^{\rm ce'}_{\rm su} \sin^2 i \simeq
1.6 \cdot 10^{11}$ yr, with corresponding maximum companion masses of
$19 M_{\rm J}$ and $2.1 M_{\rm J}$, respectively (See Fig. 2).

% Begin Tau Boo

{\bf $\tau$ Boo}. The most interesting result from our models of
$\tau$ Boo is the likely absence of a convective envelope.  We find
convective envelopes only in models with $T_{\rm eff} \lo 6350$ K
or [Fe/H]$\lo 0.07$.  Even in these
models, the convective envelope is extremely thin.  The spectral line
bisectors seen by Hatzes et al.~(1997) and others display a slight
curvature, consistent with a very thin convective layer, which our
models might not capture and which would not be significant for tidal
dissipation. Even if we include our models with the most significant
convective envelopes ($T_{\rm eff}=6200$ K, [Fe/H]=0.16), we can place an
upper limit on the size of the convective zone of $M_{ce} \lo 0.003
M_{\odot}$ (See Table 3).

While this result does not affect the possibility of tidal dissipation
{\em in the planet} circularizing the orbit as discussed in Rasio et
al.~(1996) and Rasio \& Ford (1996), it does have significant
implications for the spin history of $\tau$ Boo.  Marcy et al.~(1997)
and Drake et al.~(1998) suggest that the planet is likely to have
tidally spun up the star.  Our models indicate that $\tau$ Boo does
not have a convective envelope and thus would not have been spun up by
a close companion.  In fact, a rotation period of $3.3$ d is entirely
consistent the observed normal rotation rates of young F7 stars with
fully radiative envelopes (Gray \& Nagor 1985).  Thus, the
approximate equality of the observed rotation period and the orbital
period is most likely a coincidence and not the result of tidal
spin-up.

If we insist that tidal dissipation synchronized the stellar rotation
period, then it must have occurred much earlier in $\tau$ Boo's life,
when the star had a large convective envelope.  Our models indicate
that $\tau$ Boo's convective envelope starts to shrink rapidly near
$\sim 2\cdot 10^6$ yr and is reduced to $\lo 4 \cdot 10^{-3}
M_{\odot}$ after $\sim 1.5 \cdot 10^{7}$ yr.  For a short time, $\sim
2-3 \cdot 10^7$ yr, the synchronization timescale is $\sim 2 \cdot
10^6$ yr, providing a window of opportunity for tidal synchronization.
By $\sim 3 \cdot 10^{7}$ yr, the convective envelope has shrunk to
less than $10^{-4} M_{\odot}$.  Thus, tidal synchronization would
require that the planet be already formed and in a short-period orbit
before this time.  Indeed, it has been suggested that the planet may
have migrated to its present orbit and stopped at an orbital period
synchronized with the stellar rotation, because of the tidal force
exerted on the orbit (Lin et al.~1996; Trilling et al.~1998).  Since
the system is quite young, it is possible that the star has not had
enough time to spin down significantly after the disappearance of the
convective envelope.

% Begin Ups And 

{\bf $\upsilon$ And}. There is no measured rotation period for
$\upsilon$ And.  Although most of our models of $\upsilon$ And do have
a small convective envelope, some of the models we constructed for
$T_{\rm eff} = 6350 K$ do not.  If we adopt our best model,
 we find that the constraint $\tau^{*}_{\rm su} < \tau_{\rm
MS}$ would impose a limit of $43 M_{\rm J}$ on the companion mass,
while ignoring the reduction in efficiency from the rapid pumping
period would impose a limit of $15 M_{\rm J}$.  The constraint
that $\tau^{\rm ce}_{\rm su} < \tau_{\rm MS}$ would impose
limits of $5.5 M_{\rm J}$ for $f'=1$ and $1.9 M_{\rm J}$ for
$f=1$.

The latest radial velocity data indicate that the planet around
$\upsilon$ And may have a significantly eccentric orbit, $e \simeq
0.1$ (Marcy 1998).  If this eccentricity is confirmed, then the
non-circular orbit, combined with the survival of the planet, could
place constraints on the planet's structure.  Following Rasio et
al.~(1996) we calculate the timescales for synchronization of the
planet's rotation and for orbital circularization.  Assuming that the
planet is identical to Jupiter, we find that the rotation of the
planet should be synchronized with the orbit and that the
circularization timescales is comparable to the age of the system.
Since both these timescale are very sensitive to the radius of the
planet, we can turn the observed non-zero eccentricity into an upper
limit on the planet's radius.  Taking the age to be 2.6 Gyr and still
assuming the same dissipation factor, $Q$, as Jupiter, we find $R_{\rm
p} \lo 1.4 R_{J}$.  Unfortunately, the uncertainty in the age of the
system, and even more in $Q$, prevents this from being a stringent
constraint at this time.

% Begin Rho Cnc

{\bf $\rho^1$ Cnc}: We were not able to construct any models of
$\rho^1$ Cnc with a temperature as low as that observed by Gonzalez
(1998ab).  One possible explanation is that the $\rho^1$ Cnc system
might be an unresolved stellar binary viewed nearly face-on (Gonzalez
1998ab).

Since the star has an observed rotation period of $42$ d, which is
longer than the orbital period, we can calculate the maximum companion
mass which would not have synchronized the stellar rotation period.
We calculate $\tau^{*}_{su} \sin^2 i \simeq 1.4 \cdot 10^{15}$ yr.
Comparing this to the age of $\rho^1$ Cnc, we get a maximum companion
mass of $\sim 0.33 M_{\odot}$.  If we set $f=1$, ignoring the
reduction in efficiency from the rapid pumping period, then we obtain
$\tau^{*'}_{\rm su} \sin^2 i \simeq 1.1 \cdot 10^{14}$ yr, which
yields a maximum mass of $91 M_{\rm J}$.  We also calculate
$\tau^{\rm ce}_{\rm su} \sin^2 i \simeq 3.5 \cdot 10^{14}$ yr
and $\tau^{\rm ce'}_{\rm su} \sin^2 i \simeq 2.5 \cdot
10^{13}$ yr, which give maximum companion masses of $\sim 0.17
M_{\odot}$ and $46 M_{\rm J}$, respectively (See Fig.~4).  While
this excludes the possibility of most stellar mass companions, we can
not completely exclude the possibility of a nearly face-on stellar
companion of mass $\lo 0.3 M_{\odot}$ (Gonzalez 1998ab).
Additionally, since the companion is very close to $\rho^1$ Cnc, it
may be contaminating the spectra and distorting the observed
parameters sufficiently to affect the maximum mass we calculated here.

{\bf $\rho$ CrB}. While tidal dissipation in the planets could
circularize the orbits in 51 Peg, $\tau$ Boo, $\upsilon$ And, and
possibly also $\rho^1$ Cnc, the planet around $\rho$ CrB has too large
a semi-major axis for tidal dissipation to be significant anywhere.
Indeed, the orbit is observed to have an eccentricity of $0.15\pm0.03$
(Marcy 1998).  Depending on the orbital parameters of the secondary
star, secular perturbations may be able to produce the observed
eccentricity in the orbit of the lower mass companion.  Alternatively
this eccentricity may have been produced by a dynamical interaction
with another planet.  Alternatively, an interaction with a
circumstellar disk could induce an eccentricity, but only if the
planet is significantly more massive than its minimum mass (Artymowicz
1992).

\subsection{Metallicity Enhancements}

\subsubsection{Observational Evidence}

Observed metallicities are listed in Table 2.  Of the five short
period planets, all but $\rho$ CrB have [Fe/H]$> 0.15$.  These five
stars have a mean metallicity of $+0.18$.  If we consider the four
stars with planets in the shortest period orbits, then the average
metallicity increases to $+0.29$.  Additionally, six of the seven
stars known to have companions with $m \sin i \lo 5 M_{\rm J}$ are
metal rich relative to the sun and they have a mean metallicity of
$+0.14$. These observations have led to speculations that there may be
a relationship between stars with higher metallicities and stars with
planets.

When analyzing these metallicities, one must be careful to consider
the survey population from which the stars were selected.  As an
example, consider the SFSU planet search, which has monitored 120
stars. Of those, 67 stars are also in the Cayrel de Strobel et
al.~(1997) catalogue.  Looking at the intersection of these two lists,
we find that only 8\% have an average [Fe/H] $> 0.15.$\footnote{Cayrel
de Strobel et al.~(1997) catalogues observations of [Fe/H] from the
literature.  When the catalogue lists multiple observations for a
single star, we take the unweighted average.}  Since all the stars
known to harbor planets are F and G stars and it is more difficult to
obtain accurate metallicities for K and M stars, it may be better to
restrict our attention to the 61 F and G stars in the SFSU planet
search.  Of those, 47 appear in the Cayrel de Strobel et al.~(1997)
catalogue, and only about 5\% have an average [Fe/H] $> 0.15$.  Both
populations have a mean [Fe/H]$=-0.09$ with a standard deviation of
$0.22$.  Additional evidence for the a relationship between high
metallicities and extrasolar planets comes from a comparison with the
recent, nearly volume-limited metallicity survey of nearby solar-type
stars by Favata et al.~(1997).  They find a mean [Fe/H] of
$-0.07\pm0.26$ for their full sample and $-0.12\pm0.27$ for the
portion of the sample with $T_{\rm eff} > 5000$ K.  After they
attempt to correct for various biases, they find a peak at [Fe/H]
$\simeq -0.23$ (Favata, Micela, \& Sciortino 1997).  Extrasolar
planets in wide binary systems for which the metallicity of each star
can be accurately measured could provide useful information about the
relationship between metallicity enhancements and planet formation.
In the cases of $\tau$ Boo, $\rho^1$ Cnc, $\upsilon$ And, and $\rho$
CrB accurate determinations of the metallicity of the binary companion
are not available.  In the 16 Cygni system, spectroscopic analyses
have revealed very similar metallicities for the two stars, but slight
differences in the lithium and beryllium abundances (Gonzalez 1998;
Lopez \& Taoro 1998).  Since both these elements are involved in
nuclear reactions, complications such as rotation can significantly
affect the observed abundances.  Thus, we cannot conclude that there
is a significant difference without more careful study.  As more
extrasolar planets are discovered in wide binary systems, these
comparisons may become very instructive.

Several factors make some stars better candidates for planet
detections than others and introduce additional observational biases.
For example, high chromospheric activity increases the scatter in the
radial velocity measurements, decreasing the sensitivity of the planet
search.  Some planet searches intentionally exclude such stars, while
others observe them in spite of the reduced detection efficiency.  In
either case, this introduces complicated selection effects.  While the
selection effects are not fully understood, it seems that high stellar
metallicities may be correlated with the existence of planets, or at
least with short period planets.

\subsubsection{Mechanisms to Accrete High-Z Material}

Several mechanisms that have been discussed in the context of
extrasolar planet formation may lead to the accretion of high-Z
material onto the parent star.  While such theories might also explain
the apparent correlation of short period planets with higher
metallicities, the plausibility of such an explanation depends on many
factors.

One factor is the type of material accreted onto the star.  Accreting
material from a gaseous disk will have little effect, since this
material has essentially the same metallicity as material which formed
the star.  However, accreting a large number of terrestrial planets,
asteroids, or planetesimals could produce a noticeable effect.
Alternatively, accreting gaseous giant planets like Jupiter could also
have a significant effect since these planets are thought to contain
up to $\sim 10 M_{\oplus}$ of rocky material in their core.

The size of the stellar convective envelope at the time of accretion
is also of critical importance, since the accreted material is rapidly
diluted over the entire stellar convective region.  In particular, if
the accretion of high-Z material occurred while the star was still on
the PMS and had a large convective envelope, then there would be
little effect on the stars' observed metallicities.  To be efficient
in increasing the surface metallicity, the accretion must take place
sufficiently late in the stellar evolution when the outer convective
envelope is shallow (See Fig.~5).

Finally, the particular type of accretion process which adds high-Z
material to the star could also be important.  If small amounts of
high-Z material are gradually deposited from nearly circular orbits,
then it might remain in the outer convective envelope.  However, if a
massive object entered the star from a highly eccentric orbit, it
might penetrate more deeply into the star, thereby diluting the high-Z
material over a larger mass fraction.  Furthermore, if a rocky body is
added to the star from a nearly circular orbit, it still might sink to a
significant depth before being disrupted.  Even if the high-Z material
is not immediately mixed deep into the stellar interior, a layer of
high-Z material would have a greater mean molecular weight, which could
drive a thermal instability allowing it to penetrate deeper into the
star.  It is beyond the scope of this paper to take all of these process into
account.  Instead, we simply obtain an {\em upper bound} on the
effects of adding high-Z material to the star by noting that any
material added must have been diluted at least across the present
convective zone.

Our models show that, at present, 51 Peg, $\rho^1$ Cnc, and $\rho$ CrB
all have significant convective envelopes, while $\upsilon$ And and
$\tau$ Boo have either very thin convective envelopes or none at all
(See Table 3).  However, our best models of $\upsilon$ And and $\tau$
Boo indicate that they did have convective envelopes at the ZAMS (See
Table 3).  Since the size of the current convective envelope imposes
an upper limit on the effects of chemical pollution from accretion
independent of when the accretion occurred, we will only consider 51
Peg, $\rho^1$ Cnc, and $\rho$ CrB.  We discuss the effects of adding
various high-Z materials to these stars, but with an original
metallicity of [Fe/H]$\simeq 0$.  This metallicity is already higher
than most nearby solar-type stars, as discussed above.

{\bf Gas Giants}. Lin et al.~(1996) proposed a model in which several
multiple Jupiter-mass planets migrate inwards as a result of
dissipation in a massive disk.  Today, we are only able to observe the
last of the planets which were fortunate enough to barely avoid
spiraling into their parent star.  We consider the effect of adding
Jupiter-like giant planets with $10 M_{\oplus}$ cores of roughly
chondritic composition (Anders \& Grevesse 1989).  Starting with a solar
metallicity star and the presently observed convective envelope, it
would take $\sim 5$ and $\sim 15$ such Jupiter-like giant planets to
raise the stellar metallicities to those observed in 51 Peg and
$\rho^1$ Cnc, respectively.  However, in this model,
the accretion takes place while the star still has a gaseous
protostellar disk ($t \lo 10^7$ yr).  At this stage the star would
still be on the PMS where it would have a much larger
convective envelope.  Therefore, much more mass would be necessary to
raise their metallicities to those observed today (See Fig.~5).

In a similar model, a single massive giant planet ($\sim 4 M_{\rm
J}$) migrates towards the parent star and then undergoes Roche lobe
overflow, stabilizing its orbit against spiraling further inward
(Trilling et al.~1998).  Again, gas giant material would be accreted
onto the star, although this time the rocky core would not be
accreted.  Since Jupiter's atmosphere is only slightly enriched in
heavy elements, this is not likely to alter a star's observed
metallicity significantly.  Additionally, if the migration is due to a
viscous disk, the star will still be highly convective, rendering the
accreted material inefficient in raising the surface metallicity.

{\bf Gaseous Disks}: In both of the above models, a portion of the gas
disk can also be accreted onto the star.  The disk may have a
metallicity slightly larger than the star, but this is not expected to
significantly alter the observed stellar metallicity.  Since both the
disk mass and the disk metallicity will vary with time and neither are
well known, a quantitative estimate for this mechanism would be
difficult.

In contrast, adding rocky material (terrestrial planets, asteroids, or
planetesimals) to the star is much more efficient at increasing the
observed values of [Fe/H], as they have a much higher metallicity than
gaseous giant planets.  Given the presently observed convective
envelopes and an initially solar composition, it would require only
$\sim 40 M_{\oplus}$ and $\sim 100 M_{\oplus}$ of chondritic material
(Anders \& Grevesse 1989) to reach the observed metallicities of 51
Peg and $\rho^1$ Cnc, respectively.  However, standard planet
formation scenarios do not predict much more than $\sim 20M_{\oplus}$ in 
terrestrial planets.

{\bf Planetesimals}: Accretion of a large amount of solid material in
the form of asteroids or planetesimals arises naturally in the model
proposed by Murray et al.~(1998).  In this model giant planets can
migrate inwards while scattering planetesimals into the parent star.

Since the early migration is rapid, the planet will not clear out the
planetesimal disk until its migration slows.  Thus we neglect any
planetesimals that are scattered into the star until the planet
reaches its final orbit.  The mass accreted on the star is given by
\begin{equation}
M_{\rm acc} \simeq f(a) \frac{m}{\alpha},
\end{equation}
where 
\begin{equation}
f(a) \simeq 0.15 \left( \frac{a}{AU} \right)^{-0.374},
\end{equation}
is the fraction of planetesimals scattered onto the star by the planet
at distance $a$, $m$ is the mass of the planet, and $\alpha$ is a
parameter $\sim 0.5 - 1.0$, which must be weighted across multiple
resonances (Murray et al.~1998; Hansen 1998).  Note that $f(a)$ only
accounts for planetesimals which become planet crossing before
colliding with the star, i.e., planetesimals which collide with the
star without ever becoming planet-crossing are neglected.

Using this prescription, we can estimate the mass of planetesimals
which would be scattered into the star for 51 Peg, $\rho^1$ Cnc and
$\rho$ CrB.  We find $0.4 M_{\rm J}$ ($130 M_{\oplus}$) for 51 Peg
and $0.6 M_{\rm J}$ ($180 M_{\oplus}$) for both $\rho^1$ Cnc, and
$\rho$ CrB.  Since all this mass is in asteroids, this could lead to a
significant and observable increase in the metallicity.  Starting from
a star with solar metallicity and the present convective envelope,
adding $130 M_{\oplus}$ of asteroids to 51 Peg or $180 M_{\oplus}$ of
asteroids to $\rho^1$ Cnc increases the observed [Fe/H] to $0.48$ and
$0.39$ dex, respectively, assuming (our best model for the two stars).
However, the requirement that the planetesimal disk still be present,
implies that the stars are still young with convective envelopes
larger than at present.  The time it takes for a planet to migrate to
a 0.05 AU orbit in the Murray et al.~(1998) scenario is $\lo 3 \cdot
10^{7}$ yr, but this depends on the model of the planetesimal disk.
Once the planet stops migrating, the asteroids inside its orbit are
quickly cleared out.  As can be seen from Fig. 5, there is a window of
opportunity at $2-3 \cdot 10^{7}$ yr which is the right time to
produce the observed metallicities.  However, since the convective
envelope disappears rapidly, this requires a very specific timing.
Different models of the disk or the inclusion of additional resonances
could destroy any careful tuning of parameters in a particular model
adjusted to produce significant metallicities.

{\bf Terrestrial Planets}: Rasio \& Ford (1996) and Weidenschilling \&
Mazari (1996) proposed that 51 Peg-like planets could be produced by
gravitational scattering off a second massive planet.  If a
significant number of planets are to wind up very close to their
parent stars, a significant number should also have collided with
their parent stars.  However, it is unlikely that two planets in the
same system would be scattered so close to their parent star.  Thus,
one would not expect to see increased metallicities in stars with 51
Peg-like planets due to the accretion of giant planets.  Similarly, in
the model of Kiseleva \& Eggleton (1997), secular perturbations of a
binary companion induce large eccentricities which are later
circularized.  In either this model or the previous model, a giant
planet acquires a very high eccentricity which would disrupt inner
terrestrial planets, possibly making them collide with the parent
star. These mechanisms have the advantage that they can deposit
material onto the star after it has reached the main sequence and its
convective envelope has diminished to near its present size.  It would
take a large amount of mass in terrestrial planets and asteroids for
these mechanisms to increase the metallicity from solar to what is
observed in 51 Peg or $\rho^1$ Cnc ($\sim 40 M_{\oplus}$ and $\sim 100
M_{\oplus}$, respectively).  One would still expect the planet to
scatter asteroids into the star, as in the Murray et al.~(1998)
scenario.  However, here a particular disk mass or timing is not
required.

\subsubsection{Polluted Stellar Models}

The possibility of a star having a surface metallicity significantly
higher than its interior metallicity arises naturally in the context
of planet formation.  Thus, when constructing models based on
observational data, we should realize that the chemical composition of
the interior, which contains most of the stellar mass, could deviate
significantly from the composition observed on the stellar surface.
For this reason, it is important to consider stellar models with
interior metallicities somewhat lower than those observed.

We have constructed a few models to explore the effects of adding
high-Z material to the surface of a star.  We pick an initial
metallicity for the entire star and evolve a stellar model slightly
past the ZAMS.  Then, we increase the metallicity in the outer
convective envelope and continue to evolve the star on the main
sequence.

First, we consider the effect of adding a Jupiter with $10 M_{\oplus}$
of high-Z material (in its rocky core, using the procedure described
above) to the ZAMS Sun.  We find that the effective temperature,
luminosity, and radius increase by $ 0.55\%$, $2.6 \%$, and $0.19\%$,
respectively, while the mass of the convective envelope decreases by
approximately $5.5\%$.  Similar models have been investigated in the
context of the Solar neutrino problem and have been found to reduce
the agreement between theoretical and observational determinations of
solar p-mode frequencies (Jeffery, Mailey, \& Chambers
1997; Christensen-Dalsgaard \& Gough 1998; Bahcall 1989).

Next, we consider the consequences of a reduced interior metallicity
for our models.  Since we use observations of the luminosity,
effective temperature, and surface metallicity, the radius will not be
affected, but the mass, age, and size of the convective envelope will
be.  Generally, decreasing the interior metallicity from the observed
surface metallicity causes the mass to decrease, but the age and the
size of the convective envelope both increase.  For example, if we
take the observed parameters for 51 Peg, but set the interior
metallicity to solar, then the we obtain in our new best model a mass
of $0.94 M_{\odot}$, an age of $10.8$ Gyr, and a convective envelope
of mass $0.026 M_{\odot}$ (see Table 3 for comparison).

Finally, we have explored the possibility that a reduced interior
metallicity could explain the difficulties encountered in constructing
models that match the observed parameters for $\rho^1$ Cnc and $\tau$
Boo.  In general, lowering the interior metallicity leads to a higher
surface temperature in our models.  Therefore, in the case of $\rho^1$
Cnc, this would make the discrepancy worse.  However, this could bring
our models of $\tau$ Boo closer to agreement with the observed
effective temperature.  If we demand that our models of $\tau$ Boo
have $T_{\rm eff}=6550$ K, as indicated by the recent observations
of Gonzalez (1998a), then the difference between the interior and
surface metallicities must be $\sim 0.35$ dex.  Although possibly a
coincidence, this is comparable to $\tau$ Boo's observed
[Fe/H]=$0.34\pm0.09$.  However, since we were able to construct models
with a single metallicity for $T_{\rm eff}=6500$ K, the
uncertainty in the measurement of the effective temperature ($\simeq
100$ K) prevents us from placing a firm constraint on the interior
metallicity.  If we use the same parameters for $\tau$ Boo as
discussed in \S 3.2, but with a solar metallicity interior, we find
a mass of $1.29 M_{\odot}$, an age of $0.4$ Gyr, and no significant
outer convective envelope.

We find that a reduced interior metallicity could resolve the
difficulty in constructing models that match the observed luminosity,
temperature, and metallicity for $\tau$ Boo.  In our calculations we
have instantaneously distributed the accreted material across the
convective envelope of the star and then evolved the star in this dual
composition state.  If the material is deposited on a dynamical
timescale, then the star will temporarily be out of thermal
equilibrium.  A future calculation could evolve the star on a thermal
timescale to follow the readjustment to a new thermal equilibrium.  In
addition, if the convective envelope has a higher metallicity than the
interior, then the possibility of triggering a thermal instability
that could drive high metallicity material into the stellar interior
should be examined.

\section{Summary}

We have computed stellar models for nearby stars with close
planetary-mass companions.  Tables 5, 7, 9, 11, and 13 show the properties
of models constructed for different values of the observed parameters
within measurement uncertainties.  These models were used to calculate
the stellar parameters and error bars summarized in Table 3. We will
make a larger grid of stellar models available electronically to help
interpret future stellar observations.

Using conventional tidal dissipation theory, we have studied tidal
dissipation effects in these systems, based on the results of our
models.  We find that the orbital decay timescale is longer than the
age of the star in all systems (See Figs.~1 and 3).  The timescale for
orbital circularization is shorter than the stellar ages in 51 Peg,
$\tau$ Boo, and $\upsilon$ And, comparable to the stellar age in
$\rho^1$ Cnc, but longer than the stellar age in $\rho$ CrB.  We have
also calculated maximum companion masses based on the lack of
synchronization between the stellar rotation and the companion orbital
period for 51 Peg, $\upsilon$ And and $\rho^1$ Cnc (See Figs.~2 and
4).  For the most optimistic assumptions for the tidal coupling (in
particular assuming that the tidal torque is exerted only on the outer
convective envelope), we find $m \lo 2.1 M_{\rm J}$, 1.9
$M_{\rm J}$ and 46 $M_{\rm J}$ for 51 Peg, $\upsilon$ And, and
$\rho^1$ Cnc, respectively.  For $\tau$ Boo the similarity between the
stellar rotation period and the orbital period is likely a coincidence
and not the result of tidal spin-up.  If the significant eccentricity
of the $\upsilon$ And system is confirmed, then we can place an upper
limit on the radius of the planet, $R_p \lo 1.4 R_{\rm Jup}$.

We have examined the observational evidence for a correlation between
the presence of a planet and a high stellar metallicity.  We have
evaluated the effects of accreting high-metallicity material onto a
star in the context of various proposed mechanisms for producing short
period systems.  The accretion of gas giants or of material from a
gaseous disk causes only minor metallicity enhancements if the stellar
convective envelope is still large at the time of the
accretion. Accretion of rocky material, such as planetesimals,
asteroids, and terrestrial planets, could cause significant increases
in observed metallicities if enough mass is available.  An increased
surface metallicity can also have a significant effect on stellar
models and on the interpretation of observational data.

\acknowledgments

We are very grateful to Guillermo Gonzalez for several valuable
discussions, reanalysis of his data, and for sharing papers in advance
of publication.  We thank Dimitar Sasselov for assistance in
interpreting observational data.  We also thank Brad Hansen, Matt
Holman, and Norm Murray for their help in forming the estimates for
their model in \S 4.2, and Sallie Baliunas, Robert Donahue, Scott
Horner, and Geoff Marcy for useful comments.
This work was supported by NSF Grant AST-9618116.
E.B.F.\ is supported in part by the Orloff UROP Fund and the UROP program at MIT.
F.A.R.\ is supported in part by an Alfred P.\ Sloan Research Fellowship.
A.S. is supported in part by the Natural Sciences and Engineering Research Council of Canada.
This research has made use of the SIMBAD database, operated at CDS, Strasbourg, France.

\newpage

\figcaption[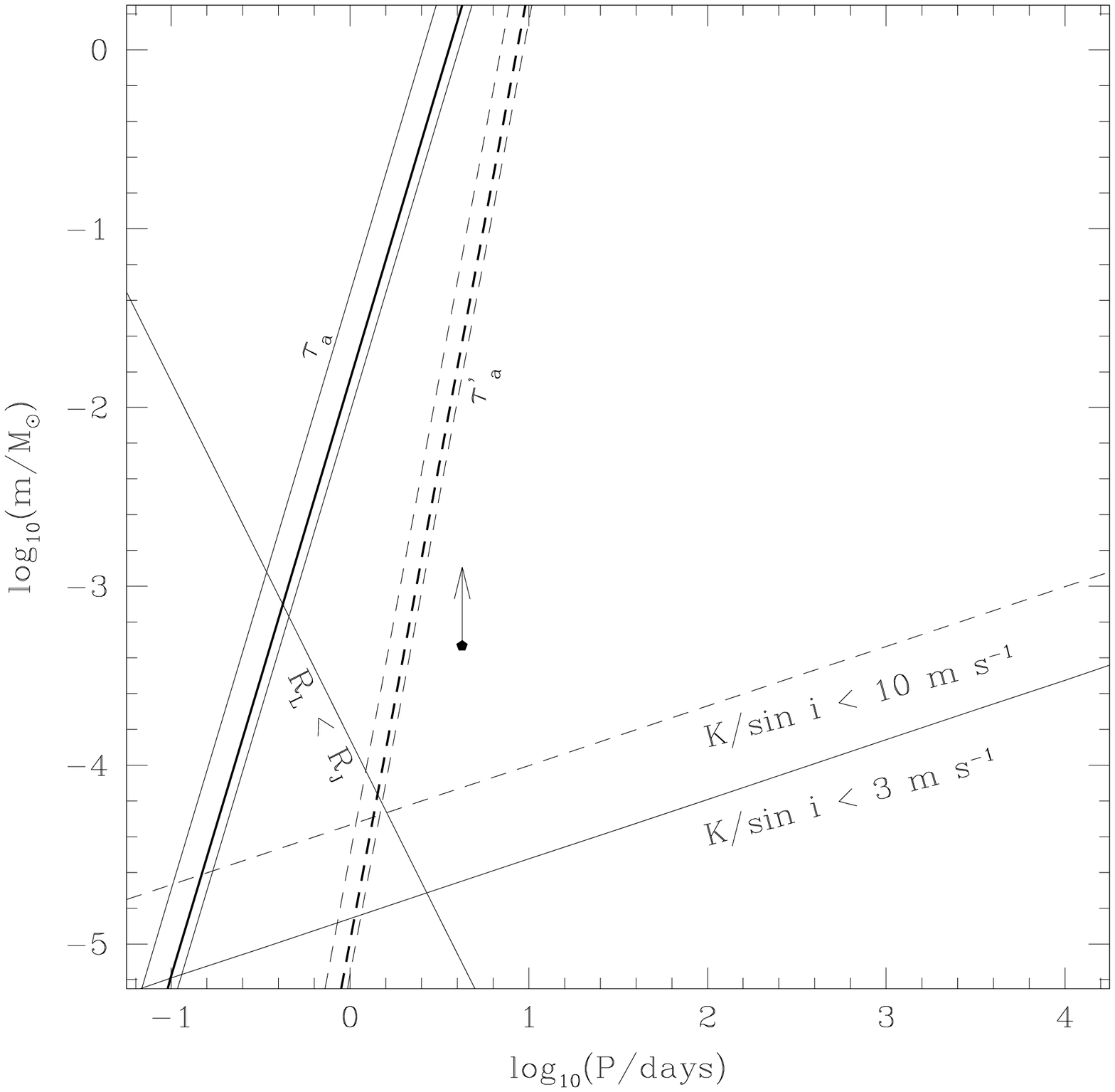]{This plot of planet mass versus
orbital period indicates the regions allowed after imposing tidal
stability constraints on 51 Pegasi.  The lines labeled $\tau_a$
indicate where the timescale for orbital decay of the planet would be
less than the age of the star, according to standard tidal dissipation
theory.  The lines labeled $\tau'_a$ differ in that no reduction of
the orbital decay timescale was applied to account for the long
convective turnover timescale (Eq.~4).  For both groups of lines
labeled $\tau_a$ and $\tau'_a$, the thicker line corresponds to our
best model and the lines on either side reflect the uncertainty in the
stellar age from our models (see Table 15). The lines labeled $K/\sin
i$ indicate the typical detection limits of current radial velocity
surveys.  The line labeled $R_L/R_J$ indicates where a planet with
mass $0.4 M_{\rm J}$, and radius $\simeq 1 R_{J} \simeq 0.1
R_{\odot}$, would overflow its Roche lobe.  The large dot indicates
the minimum mass of the planet.  This plot was constructed using our
best model parameters: $M_{*}=1.05 M_{\odot}$, $R_{*}=1.16 R_{\odot}$,
$\tau_c=18.6$ d, $M_{\rm ce}=0.023 M_{\odot}$, and $\tau_{\rm
MS}=7.6^{+4.0}_{-5.1}$ Gyr.}

\figcaption[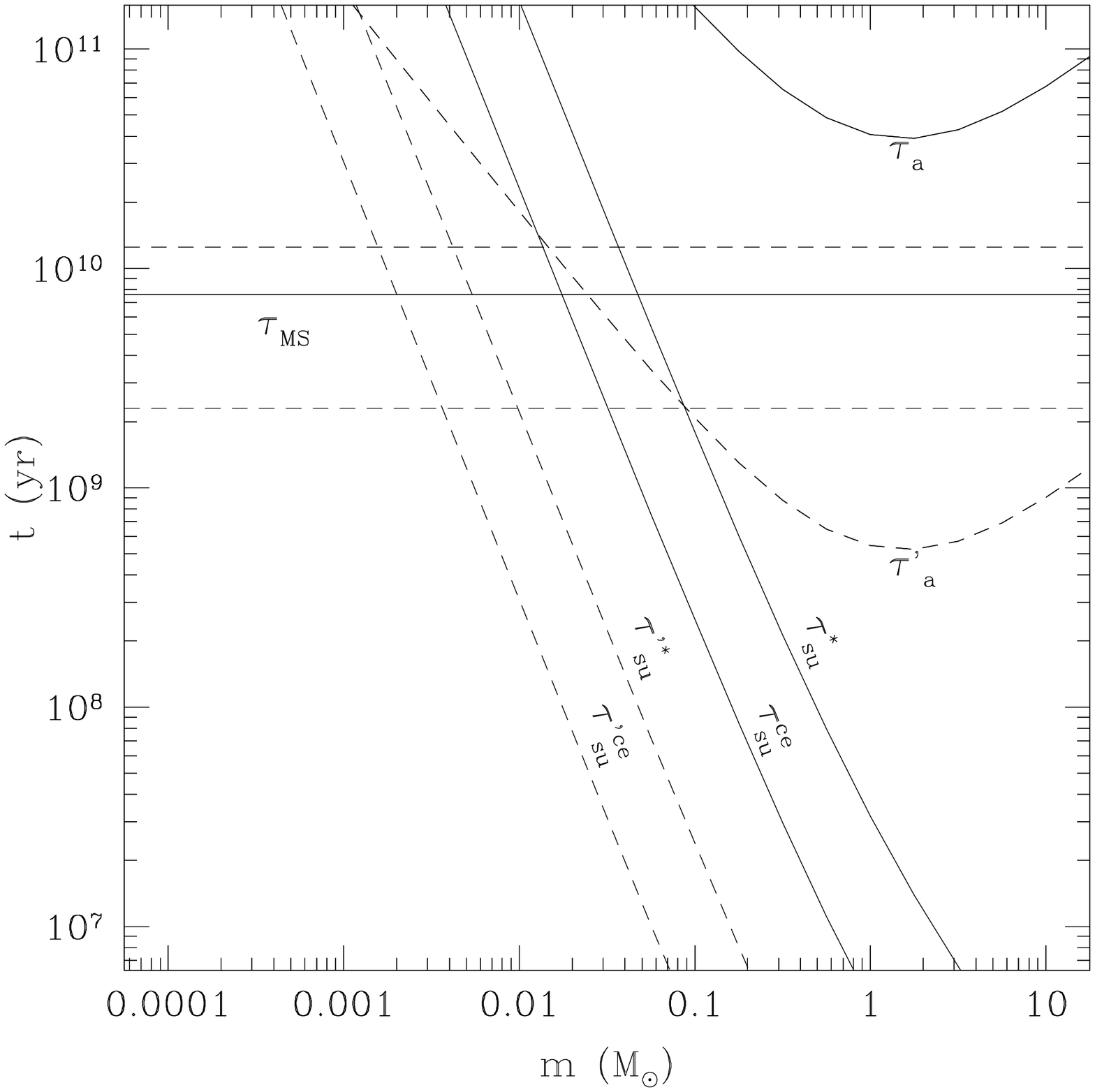]{This plot shows several time scales as a
function of the mass of the companion orbiting 51 Peg. Here
$\tau_{\rm a}$ is the orbital decay timescale, $\tau^{*}_{\rm
su}$ is the spin-up timescale for the whole star, and $\tau^{\rm
ce}_{\rm su}$ is the spin-up timescale for the convective envelope
alone.  The dashed diagonal lines correspond to setting $f=1$, while
the solid diagonal lines correspond to calculating $f$ from Eq.~4 with
$f'=1$.  The solid horizontal line indicates the age of the system,
$\tau_{\rm MS}$, as determined from our best model, while the
dashed horizontal lines indicate the uncertainty in the age.  The
intersections of the age with the spin-up timescales gives the maximum
companion masses quoted in the text.}

\figcaption[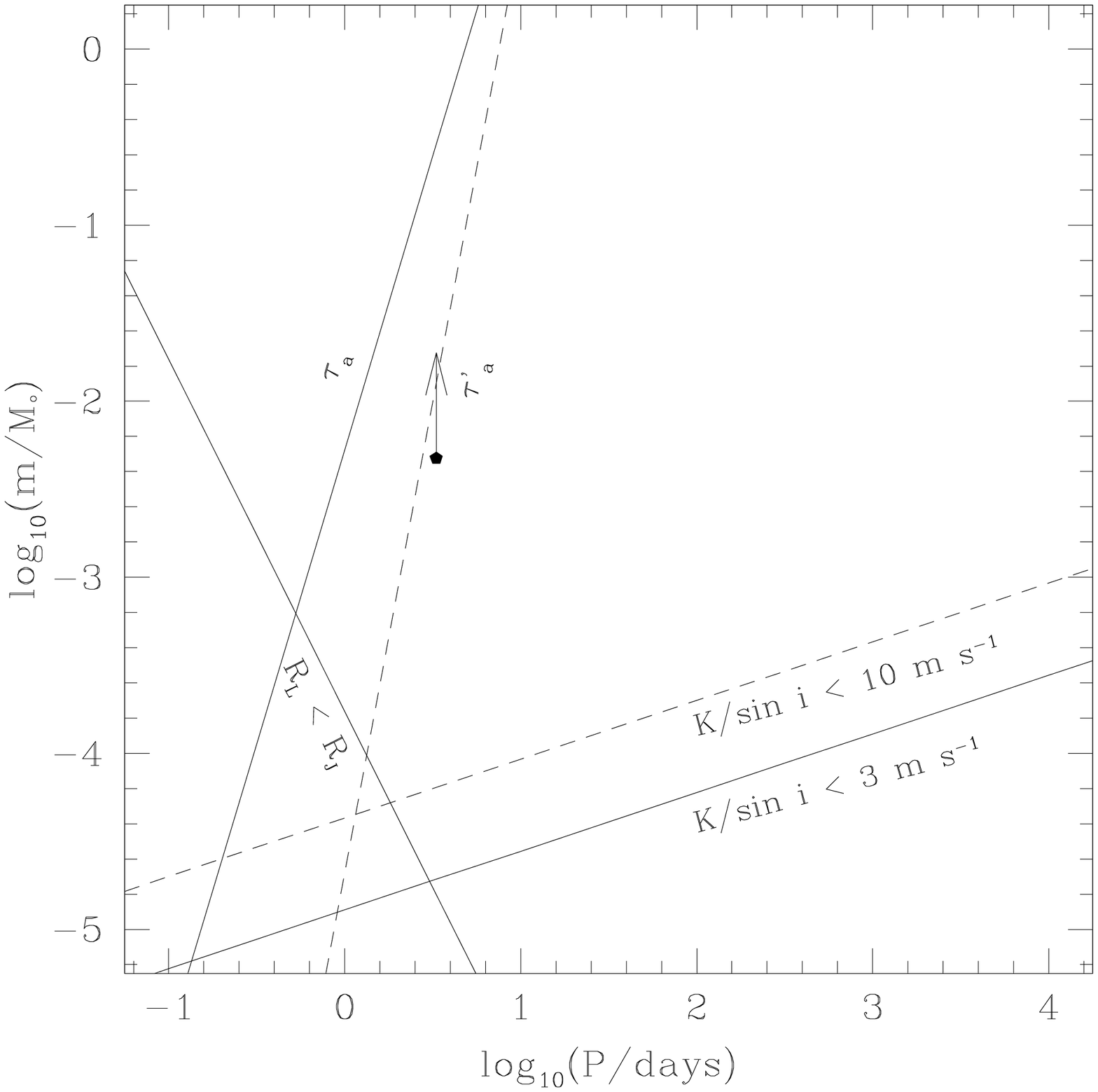]{This plot of planet mass versus
orbital period indicates the regions allowed by tidal stability
constraints on $\tau$ Boo.  Here we have taken our model {\em with the
largest convective envelope}.  The model has $M_{*}=1.30 M_{\odot}$,
$R_{*}=1.51 R_{\odot}$, $\tau_c=7.93$ d, $M_{\rm ce}=0.0035
M_{\odot}$, and an age of $2.9$ Gyr.  This model is for
$T_{\rm{eff}}=6200$ K, [Fe/H]$=0.16$, and $L_{*}=3.006 L_{\odot}$.
Other conventions are as in Fig.~1.}

\figcaption[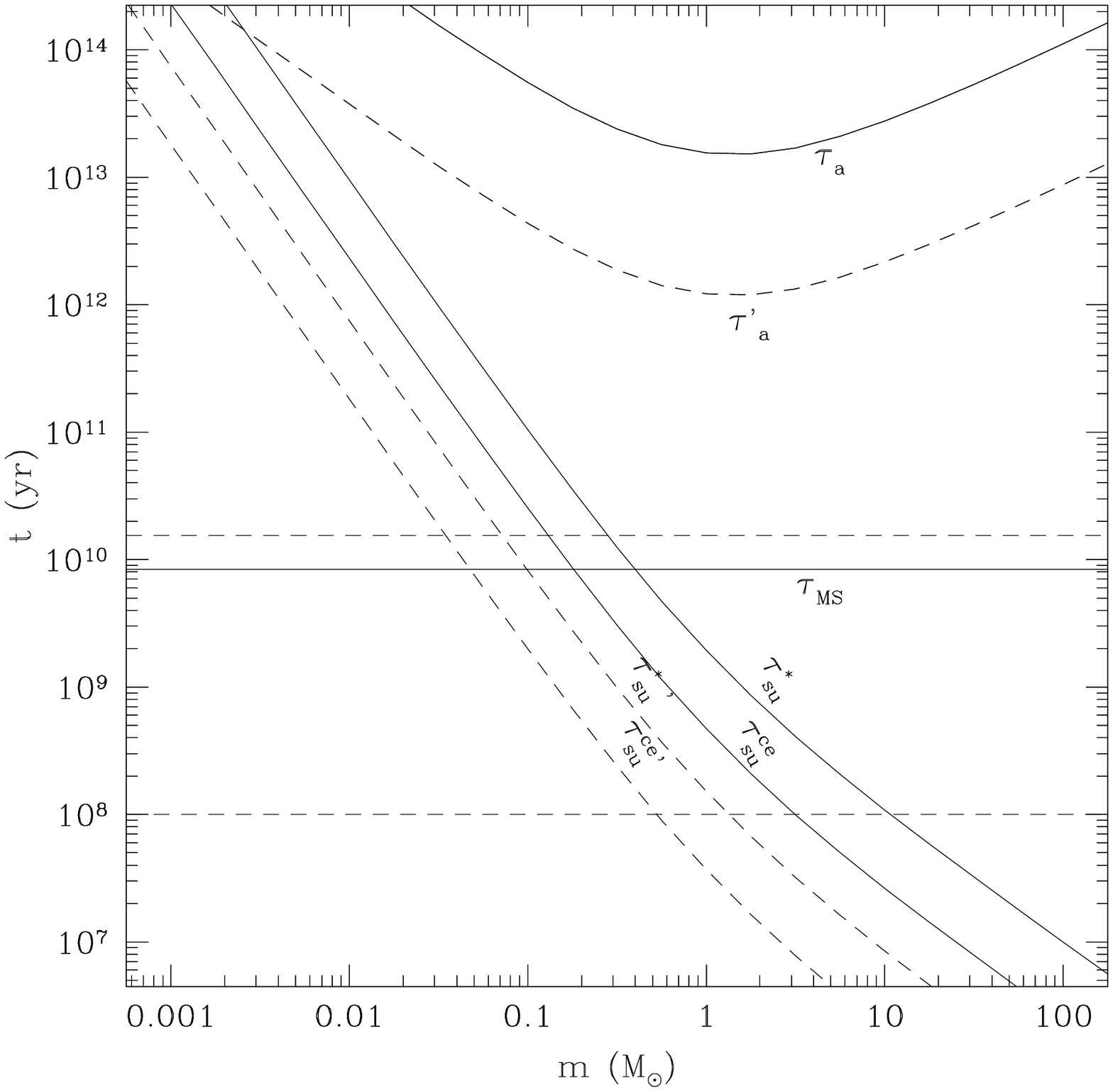]{This plot shows several time scales as
a function of the mass of the companion orbiting $\rho^1$ Cnc.  The
labels are the same as in Fig.~2.  The intersection of the age with
the timescales for tidal spin-up lead to the maximum companion masses
listed in the text.}

\figcaption[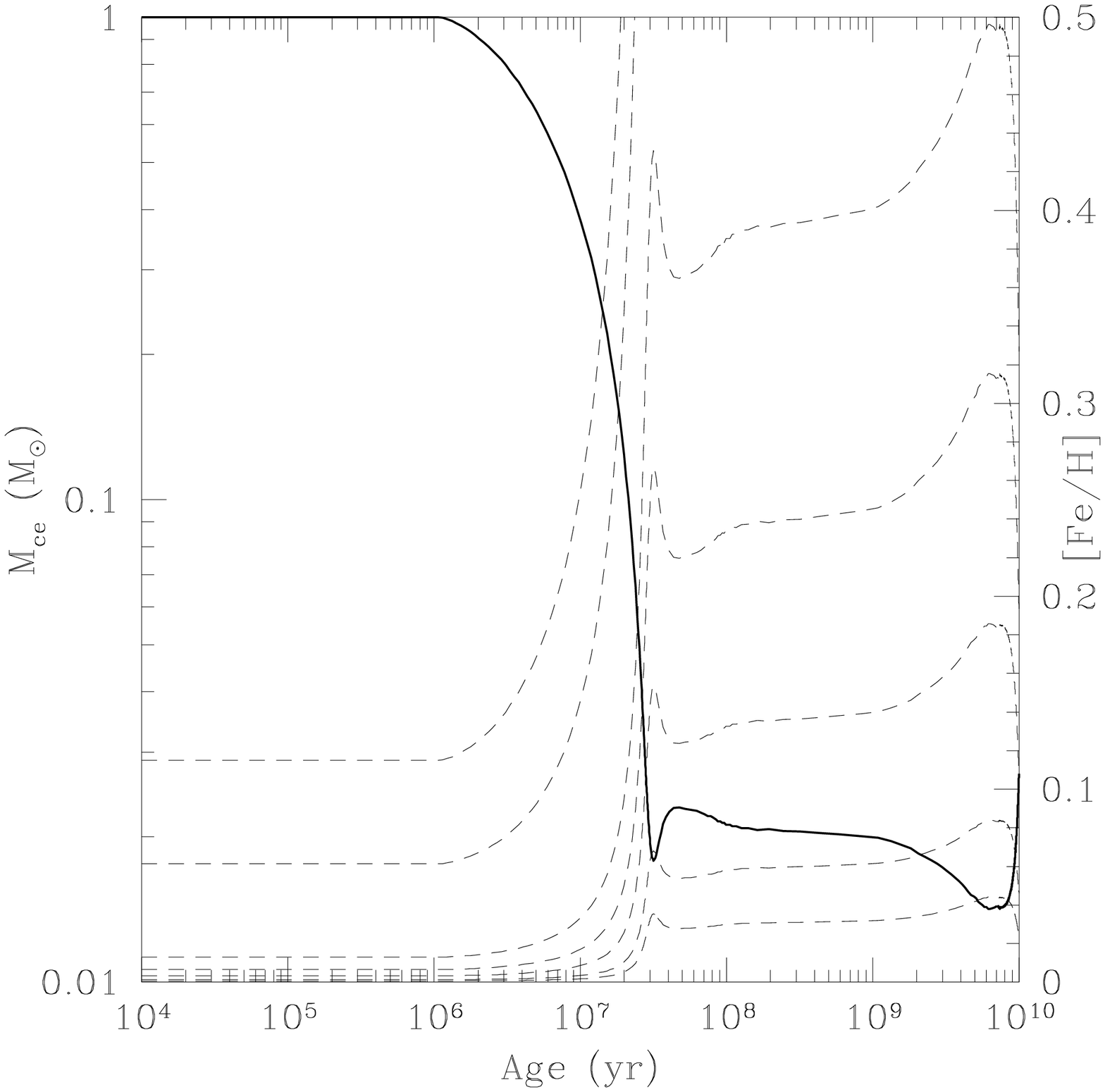]{The solid line (left axis) shows the
evolution of the Sun's convective envelope.  The dashed lines (right
axis) indicate the surface metallicity which would result from the
instantaneous accretion of rocky material onto the star at each time
in the Sun's past, assuming that the accreted material is mixed across
the convective envelope.  Starting from the bottom, the dashed lines
correspond to accreting 5, 10, 25, 50, 100, 500, and 1000
$M_{\oplus}$.  We see that producing the high surface metallicities
seen in 51 Peg (0.21 dex) and $\rho^1$ Cnc (0.29 dex) is possible with
the accretion of a large amount of rocky material after $\sim 10^7$
yr.}

\newpage

\begin{deluxetable}{cccccccc}
\tablecolumns{8}
\tablewidth{0pc}
\tablecaption{Dynamical Properties of Extrasolar Planets \tablenotemark{a}}
\scriptsize
\tablehead{ \colhead{Star} & \colhead{HD} & \colhead{$M_{*}$} 
   & \colhead{$P_{\rm orb}$} & \colhead{e}
   & \colhead{$a$} & \colhead{$m \sin i$}
   & \colhead{Ref}  \nl
   \colhead{} & \colhead{} & \colhead{$(M_{\odot})$} & \colhead{(days)} & \colhead{}
   & \colhead{(AU)} & \colhead{($M_{\rm J}$)} & \colhead{} \nl }
\startdata
$\tau$ Boo&	120136&	$1.37\pm0.09$&		$3.3128\pm0.0002$&	$0.018\pm0.016$&	$0.0413\pm0.0010$&	$4.74\pm0.10$&		[1] \nl
51 Peg&		217014&	$1.05\pm0.09$&		$4.2311\pm0.0005$&	$0.012\pm0.010$&		$0.0520\pm0.0015$&	$0.464\pm0.036$&	[2] \nl
$\upsilon$ And&	9826&	$1.34\pm0.12$&		$4.611\pm0.005$&	$0.109\pm0.040$&		$0.0598\pm0.018$&	$0.739\pm0.075$&	[1] \nl
$\rho^1$ Cnc&	75732&	$0.95\pm0.10$&		$14.648\pm0.0009$&	$0.051\pm0.013$&	$0.1152\pm0.0038$&	$0.874\pm0.079$&	[1] \nl
$\rho$ CrB&	143761&	$0.89\pm0.05$&		$39.645\pm0.088$&	$0.15\pm0.03$&		$0.2192\pm0.0042$&	$1.064\pm0.059$&	[3,7] \nl
HD 114762&	114762&	$0.75\pm0.15$&		$83.90\pm0.08$&		$0.35\pm0.05$&		$0.34\pm0.04$&		$10.5\pm1.9$&		[4] \nl
70 Vir&		117176&	$1.01^{+0.05}_{-0.02}$&	$116.67\pm0.01$&	$0.4\pm0.01$&		$0.47\pm0.01$&		$7.3\pm0.3$&		[5] \nl
16 Cyg B&	186427&	$  0.96\pm0.05$&	$800.8\pm11.7$&		$0.634\pm0.082$&	$1.66\pm0.05$&		$2.0\pm0.3$&		[6] \nl
47 UMa&		95128&	$  1.01\pm0.05$&	$1090\pm15$&		$0.03\pm0.06$&		$2.08\pm0.06$&		$2.3\pm0.2$&		[5] \nl
\enddata
\normalsize

\tablenotetext{a}{The orbital parameters (orbital period $P_{\rm
orb}$, eccentricity $e$, semimajor axis $a$, and minimum mass $m \sin
i$) are from: [1] Butler \& Marcy 1996; [2] Marcy et al. 1997; [3]
Noyes et al. 1997; [4] Mazeh et al. 1996; [5] Marcy \& Butler 1996;
[6] Cochran et al. 1997a; [7] Marcy 1998. In order to determine $a$ and
$m \sin i$, we have assumed the stellar mass given in column 3.  The
first five stellar masses are based on the results of this paper and
are discussed later.  The remaining stellar masses are based on
preliminary calculations which will be presented in a subsequent
paper.
}

\normalsize
\end{deluxetable}

\begin{deluxetable}{cccccccc}
\scriptsize
\tablecolumns{8}
\tablewidth{0pc}

\tablecaption{Observed Parameters of Stars with Extrasolar Planets \tablenotemark{b}}

\tablehead{ \colhead{Star} & \colhead{MK} & \colhead{$M_V$} & \colhead{$T_{\rm eff}$ (K)} &
   \colhead{[Fe/H]} & \colhead{log g (cgs)} & \colhead{ BC (V)} &
   \colhead{ $L_{*}/L_{\odot}$} }

\startdata
$\tau$ Boo&		F7V&	$3.535\pm0.024$&	$6550\pm100$&	$+0.34\pm0.09$&	$4.50\pm0.15$&	+0.02&	$3.006\pm0.213$ \nl
51 Peg&			G5V&	$4.518\pm0.025$&	$5750\pm75$&	$+0.21\pm0.06$&	$4.40\pm0.10$&	-0.07&	$1.321\pm0.094$ \nl
$\upsilon$ And&		F8V&	$3.453\pm0.021$&	$6250\pm100$&	$+0.17\pm0.08$&	$4.30\pm0.10$&	-0.01&	$3.333\pm0.225$ \nl
$\rho^1$ Cnc&		G8V&	$5.470\pm0.023$&	$5250\pm70$&	$+0.45\pm0.03$&	$4.40\pm0.15$&	-0.18&	$0.608\pm0.042$ \nl
$\rho$ CrB&		G2V&	$4.184\pm0.027$&	$5750\pm75$&	$-0.29\pm0.06$&	$4.10\pm0.05$&	-0.07&	$1.796\pm0.132$ \nl
HD 114762&		F9V&	$4.259\pm0.123$&	$5950\pm75$&	$-0.60\pm0.06$&	$4.45\pm0.05$&	-0.07&	$1.676\pm0.290$ \nl
70 Vir&			G5V&	$3.680\pm0.029$&	$5500\pm75$&	$-0.03\pm0.06$&	$3.90\pm0.10$&	-0.07&	$2.856\pm0.076$ \nl
16 Cyg B&		G5V&	$4.597\pm0.024$&	$5700\pm75$&	$+0.06\pm0.06$&	$4.35\pm0.05$&	-0.07&	$1.228\pm0.087$ \nl
47 UMa&			G0V&	$4.288\pm0.020$&	$5800\pm75$&	$+0.01\pm0.06$&	$4.25\pm0.05$&	-0.05&	$1.603\pm0.106$ \nl
\enddata

\tablenotetext{b}{Spectral type from Hipparcos Catalogue; $M_V$
derived from Hipparcos Catalogue (ESA, 1997) photometric and
parallax data; $T_{\rm eff}$, [Fe/H], and $\log g$ from Gonzalez
1997ab, 1998ab; BC(V) derived from previous parameters using color
calibration as in the Yale isochrones (Green et al.~1987);
$L_{*}/L_{\odot}$ calculated from the Hipparcos absolute visual
magnitude and BC(V).}

\normalsize
\end{deluxetable}

\begin{deluxetable}{cccccccccc}
\scriptsize
\tablecolumns{10}
\tablewidth{0pc}
\label{TableResults1}
\tablecaption{Primary Results of Stellar Model Calculations}

\tablehead{ \colhead{Star} & \colhead{$M_{*}$} & \colhead{$R_{*}$} & \colhead{Age} & \colhead{$M_{\rm ce}$} & \colhead{$R_{\rm ce}$} &  \colhead{$\tau_c$} &  \colhead{$k^2_{*}$}  & \colhead{$k^2_{\rm ce}$} &  \colhead{$M_{\rm ce, ZAMS}$} 
\nl
\colhead{} & \colhead{($M_{\odot}$)} & \colhead{($R_{\odot}$)} & \colhead{(GYr)} & \colhead{($M_{\odot}$)} & \colhead{($R_{\odot}$)} &  \colhead{(d)} &  \colhead{}  & \colhead{} &  \colhead{($M_{\odot}$)} }
\startdata
$\tau$ Boo& 	$1.37\pm0.08$&		$1.41^{+0.10}_{-0.09}$&	$1.2^{+1.2}_{-0.8}$&	$\lo 0.002$&			$1.19\pm0.09$&		$\lo 5.8$&		0.10&	$\lo 0.002$ 	&0.008 \nl 
\nl 
51 Peg&		$1.05^{+0.09}_{-0.08}$& $1.16\pm0.07$& 		$7.6^{+4.0}_{-5.1}$&	$0.023^{+0.007}_{-0.006}$& 	$0.82\pm0.04$&		$18.6\pm2.5$&		0.12&	0.020&		0.037 \nl 
\nl 
$\upsilon$ And&	$1.34^{+0.07}_{-0.12}$&	$1.56^{+0.11}_{-0.10}$&	$2.6^{+2.1}_{-1.0}$&	$0.002^{+0.003}_{-0.002}$&	$1.29^{+0.07}_{-0.06}$&	$6.8^{+2.3}_{-6.8}$&	0.10&	0.002&		0.006  \nl 
\nl 
$\rho^1$ Cnc&	$0.95^{+0.11}_{-0.09}$&	$0.93^{+0.02}_{-0.03}$&	$8.4^{+7.1}_{-8.3}$&	$0.046^{+0.004}_{-0.006}$&	$0.63\pm0.03$&		$26.7^{+1.2}_{-2.2}$&	0.15&	0.048&		0.069  \nl 
\nl
$\rho$ CrB&	$0.89^{+0.05}_{-0.04}$&	$1.35^{+0.09}_{-0.08}$&	$14.1^{+2.0}_{-2.4}$&	$0.033^{+0.011}_{-0.009}$&	$0.88^{+0.05}_{-0.03}$&	$21.5^{+2.9}_{-2.8}$&	0.10&	0.021&		0.029	\nl 
\nl
\enddata

\normalsize
\end{deluxetable}

\begin{deluxetable}{ccccl}
\small
\tablecaption{Summary of Observational Data for 51 Peg}
\tablecolumns{5}
\tablewidth{0pc}
\tablehead{ \colhead{$T_{\rm eff}$ (K)}& \colhead{[Fe/H] (solar)}& \colhead{$\log g$ (cgs)}& \colhead{$v \sin i$ (km s$^{-1}$)}& \colhead{Reference} }
\startdata
$5750\pm75$&	$0.22\pm 0.06$&	$4.40\pm0.10$&	$1.4\pm0.3$&	Gonzalez 1997a,~1998 \\
$5793\pm70$&	$0.20 \pm0.07$&	$4.33\pm0.10$&	$2.0\pm1.0$&	Fuhrmann et al.~1997 \\
&		&		&		$2.35\pm0.1$&	Hatzes et al.~1997 \\
$5775$&	        $0.20\pm 0.07$&	4.35&		&		Tomkin et al.~1997 \\
&		&		&		$2.4\pm0.3$&	Francois et al.~1996 \\
5669&		&		4.06&		&		Gratton et al.~1996 \\
&		$0.172\pm0.049$&&		&		Taylor 1996\tablenotemark{c} \\
&		0.19&		4.32&		$2.8\pm0.5$&	Mayor \& Queloz 1995 \\
5730&		&		&		&		Gray 1995 \\
5773&		&		&		&		Grennon (cited by Mayor \& Queloz 1995) \\
5724&		&		4.30&		&		Valenti (cited by Mayor \& Queloz 1995) \\
5755&		0.06&		4.18&		&		Edvardsson 1993b \\
5740&		0.05&		3.76&   	&		McWilliam 1990 \\
&		&		4.13&		&		Xu 1991 \\
5750&		0.12&		&		$1.7\pm0.8$&	Soderblom 1982 \\
5727&		0.12&		4.27&		&		Hearnshaw 1972 \\ 
\enddata
\tablenotetext{c}{This value is based on a statistical analysis of
previous measurements and not any new observations.}
\normalsize
\end{deluxetable}

%\documentstyle[10pt,aaspp4]{article}
%\begin{document}
\begin{deluxetable}{cccccccccccccccc}
\scriptsize
\tablecolumns{16}
\tablewidth{0pc}
\tablecaption{Stellar Models for 51 Peg}
\tablehead{ \colhead{$M_{*}$} & \colhead{ $\left[\frac{Fe}{H}\right]$ } & \colhead{ $T_{\rm eff}$ } & \colhead{ $L_{*}$} & \colhead{ Z } & \colhead{ X } & \colhead{ Y } & \colhead{ Age } & \colhead{ $M_{ce}$ } & \colhead{ PSH } & \colhead{ $R_{ce}$ } & \colhead{ $L_{*}$ } & \colhead{ $R_{*}$ } & \colhead{ $T_{\rm eff}$ } & \colhead{ $\log g$ } & \colhead{ $\frac{\delta Y}{\delta Z}$ } \nl 
\colhead{Solar} & \colhead{Solar} & \colhead{ K } & \colhead{Solar} & \colhead{ } & \colhead{ } & \colhead{ } & \colhead{Gyr} & \colhead{Solar} & \colhead{Solar} & \colhead{Solar} & \colhead{Solar} & \colhead{Solar} & \colhead{K} & \colhead{cgs} & \colhead{Solar}  } 
\startdata
1.092&	0.21&	5750&	1.321&	0.030&	0.687&	0.283&	6.4&	0.02226&	0.0942&	0.835&	1.322&	1.161&	5751&	4.35&	0.000\nl 
1.067&	0.21&	5750&	1.321&	0.030&	0.669&	0.301&	7.2&	0.02313&	0.0954&	0.827&	1.321&	1.161&	5750&	4.34&	1.500\nl 
1.050&	0.21&	5750&	1.321&	0.030&	0.657&	0.312&	7.6&	0.02371&	0.0962&	0.822&	1.321&	1.161&	5750&	4.33&	2.500\nl 
1.033&	0.21&	5750&	1.321&	0.030&	0.646&	0.324&	8.1&	0.02429&	0.0970&	0.816&	1.320&	1.161&	5749&	4.32&	3.500\nl 
\tableline
1.059&	0.21&	5675&	1.321&	0.030&	0.687&	0.283&	8.8&	0.02772&	0.1013&	0.833&	1.321&	1.192&	5675&	4.31&	0.000\nl 
1.033&	0.21&	5675&	1.321&	0.030&	0.669&	0.301&	9.5&	0.02856&	0.1024&	0.824&	1.321&	1.192&	5675&	4.30&	1.500\nl 
1.016&	0.21&	5675&	1.321&	0.030&	0.657&	0.312&	10.1&	0.02912&	0.1031&	0.819&	1.321&	1.192&	5675&	4.29&	2.500\nl 
0.998&	0.21&	5675&	1.321&	0.030&	0.646&	0.324&	10.6&	0.02968&	0.1039&	0.813&	1.321&	1.192&	5675&	4.28&	3.500\nl 
1.124&	0.21&	5825&	1.321&	0.030&	0.687&	0.283&	4.2&	0.01769&	0.0875&	0.836&	1.324&	1.132&	5828&	4.38&	0.000\nl 
1.100&	0.21&	5825&	1.321&	0.030&	0.669&	0.301&	4.9&	0.01843&	0.0886&	0.829&	1.324&	1.132&	5827&	4.37&	1.500\nl 
1.084&	0.21&	5825&	1.321&	0.030&	0.657&	0.312&	5.3&	0.01893&	0.0893&	0.824&	1.323&	1.132&	5827&	4.37&	2.500\nl 
1.068&	0.21&	5825&	1.321&	0.030&	0.646&	0.324&	5.8&	0.01942&	0.0901&	0.819&	1.323&	1.132&	5826&	4.36&	3.500\nl 
\tableline
1.050&	0.15&	5750&	1.321&	0.027&	0.690&	0.283&	7.8&	0.02328&	0.0960&	0.825&	1.325&	1.161&	5755&	4.33&	0.000\nl 
1.032&	0.15&	5750&	1.321&	0.027&	0.679&	0.295&	8.4&	0.02390&	0.0968&	0.819&	1.324&	1.161&	5753&	4.32&	1.500\nl 
1.020&	0.15&	5750&	1.321&	0.027&	0.671&	0.302&	8.8&	0.02431&	0.0974&	0.815&	1.323&	1.161&	5752&	4.32&	2.500\nl 
1.008&	0.15&	5750&	1.321&	0.027&	0.663&	0.310&	9.2&	0.02472&	0.0980&	0.811&	1.322&	1.161&	5751&	4.31&	3.500\nl 
1.099&	0.27&	5750&	1.321&	0.035&	0.658&	0.307&	6.0&	0.02224&	0.0938&	0.835&	1.321&	1.161&	5750&	4.35&	1.500\nl 
1.073&	0.27&	5750&	1.321&	0.035&	0.641&	0.324&	6.7&	0.02294&	0.0950&	0.828&	1.321&	1.161&	5750&	4.34&	2.500\nl 
1.048&	0.27&	5750&	1.321&	0.035&	0.625&	0.340&	7.5&	0.02364&	0.0962&	0.820&	1.321&	1.161&	5749&	4.33&	3.500\nl 
\tableline
1.092&	0.21&	5750&	1.233&	0.030&	0.687&	0.283&	5.6&	0.02203&	0.0902&	0.812&	1.234&	1.122&	5751&	4.38&	0.000\nl 
1.068&	0.21&	5750&	1.233&	0.030&	0.669&	0.301&	6.3&	0.02287&	0.0913&	0.804&	1.233&	1.122&	5750&	4.37&	1.500\nl 
1.052&	0.21&	5750&	1.233&	0.030&	0.657&	0.312&	6.8&	0.02343&	0.0920&	0.799&	1.233&	1.122&	5750&	4.36&	2.500\nl 
1.036&	0.21&	5750&	1.233&	0.030&	0.646&	0.324&	7.3&	0.02400&	0.0928&	0.794&	1.233&	1.122&	5750&	4.35&	3.500\nl 
1.094&	0.21&	5750&	1.415&	0.030&	0.687&	0.283&	7.1&	0.02251&	0.0982&	0.860&	1.415&	1.202&	5750&	4.32&	0.000\nl 
1.068&	0.21&	5750&	1.415&	0.030&	0.669&	0.301&	7.8&	0.02328&	0.0994&	0.852&	1.415&	1.202&	5750&	4.31&	1.500\nl 
1.050&	0.21&	5750&	1.415&	0.030&	0.657&	0.312&	8.2&	0.02379&	0.1002&	0.846&	1.415&	1.202&	5750&	4.30&	2.500\nl 
1.033&	0.21&	5750&	1.415&	0.030&	0.646&	0.324&	8.7&	0.02431&	0.1010&	0.841&	1.415&	1.202&	5750&	4.29&	3.500\nl 
\enddata
\normalsize
\end{deluxetable}
%\end{document}

\begin{deluxetable}{ccccl}
\small
\tablecaption{Summary of Observational Data for $\tau$ Boo}
\tablecolumns{5}
\tablewidth{0pc}
\tablehead{ \colhead{$T_{\rm eff}$ (K)}& \colhead{[Fe/H] (solar)}& \colhead{$\log g$ (cgs)}& \colhead{$v \sin i$ (km s$^{-1}$)}& \colhead{Reference} }
\startdata
$6360\pm80$&	$0.27 \pm0.08$&	$4.17\pm0.10$&	$15.6\pm0.7$&	Fuhrmann et al.~1998 \\
$6550\pm100$&	&		&		&		Gonzalez 1997b (spectroscopic upon additional analysis) \\
$6600\pm100$&	$0.34\pm 0.09$&	$4.50\pm0.15$&	$14.5\pm0.5$&	Gonzalez 1997a, 1998 (spectroscopic) \\
6405&		0.18&		&		&		Gonzalez 1997a (photometric) \\
&		&		&		$15\pm1$&	Baliunas et al.~1997 \\ 
6358&		0.09&		4.22&		&		Marsakov et al.~1995 \\
6390&   	0.30&		3.8&		&		Boesgaard \& Lavery~1986  \\
6460&   	0.00&		4.3&		&		Thevenin et al.~1986 \\
6300&		0.00&		4.60&		&		Thevenin \& Foy 1983 \\
&		&		&		$14.8\pm0.3$&	Gray 1982 \\
6490&		0.21&		&		$17\pm1$&	Soderblom 1982 \\
6380&   	0.14&		4.3&		&		Kuroczkin \& Wiszniewski~1977 \\
6450&		0.28&		4.3&		&		Perrin 1977 \\ 
6462&		0.28&		4.3&		&		Spite 1968 \\ \tableline

\enddata
\normalsize
\end{deluxetable}

%\documentstyle[10pt,aaspp4]{article}
%\begin{document}
\begin{deluxetable}{cccccccccccccccc}
\scriptsize
\tablecolumns{16}
\tablewidth{0pc}
\tablecaption{Stellar Models for $\tau$ Boo}
\tablehead{ \colhead{$M_{*}$} & \colhead{ $\left[\frac{Fe}{H}\right]$ } & \colhead{ $T_{\rm eff}$ } & \colhead{ $L_{*}$} & \colhead{ Z } & \colhead{ X } & \colhead{ Y } & \colhead{ Age } & \colhead{ $M_{ce}$ } & \colhead{ PSH } & \colhead{ $R_{ce}$ } & \colhead{ $L_{*}$ } & \colhead{ $R_{*}$ } & \colhead{ $T_{\rm eff}$ } & \colhead{ $\log g$ } & \colhead{ $\frac{\delta Y}{\delta Z}$ } \nl 
\colhead{Solar} & \colhead{Solar} & \colhead{ K } & \colhead{Solar} & \colhead{ } & \colhead{ } & \colhead{ } & \colhead{Gyr} & \colhead{Solar} & \colhead{Solar} & \colhead{Solar} & \colhead{Solar} & \colhead{Solar} & \colhead{K} & \colhead{cgs} & \colhead{Solar}  } 
\startdata
1.403&	0.25&	6400&	2.808&	0.033&	0.684&	0.283&	0.5&	0.00000&	0.0000&	1.154&	2.816&	1.367&	6404&	4.31&	0.000\nl 
1.375&	0.25&	6400&	2.808&	0.033&	0.662&	0.305&	0.7&	0.00000&	0.0000&	1.157&	2.812&	1.367&	6402&	4.30&	1.500\nl 
1.357&	0.25&	6400&	2.808&	0.033&	0.647&	0.320&	0.9&	0.00000&	0.0000&	1.159&	2.809&	1.367&	6401&	4.30&	2.500\nl 
1.338&	0.25&	6400&	2.808&	0.033&	0.632&	0.334&	1.1&	0.00000&	0.0000&	1.161&	2.807&	1.366&	6399&	4.29&	3.500\nl 
\tableline
1.350&	0.16&	6400&	3.006&	0.027&	0.690&	0.283&	1.5&	0.00000&	0.0000&	1.191&	3.008&	1.414&	6401&	4.27&	0.000\nl 
1.337&	0.16&	6400&	3.006&	0.027&	0.677&	0.296&	1.6&	0.00000&	0.0000&	1.193&	3.007&	1.414&	6400&	4.26&	1.500\nl 
1.329&	0.16&	6400&	3.006&	0.027&	0.669&	0.304&	1.7&	0.00000&	0.0000&	1.194&	3.007&	1.414&	6400&	4.26&	2.500\nl 
1.320&	0.16&	6400&	3.006&	0.027&	0.661&	0.312&	1.7&	0.00000&	0.0000&	1.196&	3.007&	1.414&	6400&	4.26&	3.500\nl 
1.450&	0.34&	6350&	3.006&	0.041&	0.676&	0.283&	0.8&	0.00000&	0.0000&	1.230&	3.015&	1.437&	6353&	4.28&	0.000\nl 
1.416&	0.34&	6350&	3.006&	0.041&	0.642&	0.316&	1.0&	0.00000&	0.0000&	1.224&	3.013&	1.437&	6353&	4.27&	1.500\nl 
1.392&	0.34&	6350&	3.006&	0.041&	0.620&	0.339&	1.1&	0.00024&	0.0112&	1.220&	3.012&	1.437&	6353&	4.27&	2.500\nl 
1.369&	0.34&	6350&	3.006&	0.041&	0.598&	0.361&	1.3&	0.00096&	0.0452&	1.215&	3.011&	1.436&	6352&	4.26&	3.500\nl 
\tableline
1.383&	0.25&	6300&	3.006&	0.033&	0.684&	0.283&	1.7&	0.00140&	0.0729&	1.241&	3.013&	1.460&	6302&	4.25&	0.000\nl 
1.362&	0.25&	6300&	3.006&	0.033&	0.662&	0.305&	1.8&	0.00163&	0.0750&	1.232&	3.013&	1.460&	6302&	4.24&	1.500\nl 
1.349&	0.25&	6300&	3.006&	0.033&	0.647&	0.320&	1.9&	0.00178&	0.0763&	1.227&	3.013&	1.460&	6303&	4.24&	2.500\nl 
1.335&	0.25&	6300&	3.006&	0.033&	0.632&	0.334&	2.0&	0.00193&	0.0777&	1.221&	3.013&	1.459&	6303&	4.23&	3.500\nl 
1.423&	0.25&	6450&	3.006&	0.033&	0.684&	0.283&	0.5&	0.00000&	0.0000&	1.159&	3.008&	1.392&	6451&	4.30&	0.000\nl 
1.397&	0.25&	6450&	3.006&	0.033&	0.662&	0.305&	0.7&	0.00000&	0.0000&	1.162&	3.008&	1.392&	6451&	4.30&	1.500\nl 
1.380&	0.25&	6450&	3.006&	0.033&	0.647&	0.320&	0.8&	0.00000&	0.0000&	1.164&	3.008&	1.392&	6451&	4.29&	2.500\nl 
1.363&	0.25&	6450&	3.006&	0.033&	0.632&	0.334&	1.0&	0.00000&	0.0000&	1.166&	3.008&	1.392&	6451&	4.28&	3.500\nl 
\tableline
1.408&	0.25&	6400&	3.006&	0.033&	0.684&	0.283&	0.9&	0.00000&	0.0000&	1.189&	3.005&	1.414&	6400&	4.29&	0.000\nl 
1.384&	0.25&	6400&	3.006&	0.033&	0.662&	0.305&	1.1&	0.00000&	0.0000&	1.192&	3.006&	1.414&	6400&	4.28&	1.500\nl 
1.368&	0.25&	6400&	3.006&	0.033&	0.647&	0.320&	1.2&	0.00000&	0.0000&	1.194&	3.007&	1.414&	6400&	4.27&	2.500\nl 
1.351&	0.25&	6400&	3.006&	0.033&	0.632&	0.334&	1.3&	0.00000&	0.0000&	1.196&	3.008&	1.414&	6401&	4.27&	3.500\nl 
1.419&	0.25&	6400&	3.218&	0.033&	0.684&	0.283&	1.2&	0.00000&	0.0000&	1.224&	3.220&	1.463&	6401&	4.26&	0.000\nl 
1.396&	0.25&	6400&	3.218&	0.033&	0.662&	0.305&	1.3&	0.00000&	0.0000&	1.228&	3.218&	1.463&	6400&	4.25&	1.500\nl 
1.381&	0.25&	6400&	3.218&	0.033&	0.647&	0.320&	1.4&	0.00000&	0.0000&	1.230&	3.216&	1.463&	6399&	4.25&	2.500\nl 
1.366&	0.25&	6400&	3.218&	0.033&	0.632&	0.334&	1.5&	0.00000&	0.0000&	1.233&	3.215&	1.463&	6398&	4.24&	3.500\nl 
\enddata
\normalsize
\end{deluxetable}
%\end{document}

\begin{deluxetable}{ccccl}
\small
\tablecaption{Summary of Observational Data for $\upsilon$ And}
\tablecolumns{5}
\tablewidth{0pc}
\tablehead{ \colhead{$T_{\rm eff}$ (K)}& \colhead{[Fe/H] (solar)}& \colhead{$\log g$ (cgs)}& \colhead{$v \sin i$ (km s$^{-1}$)}& \colhead{Reference} }
\startdata
$6107\pm80$&	$0.09 \pm0.06$&		$4.01\pm0.10$&	$9.5\pm0.8$&	Fuhrmann et al.~1998 \\
$6250\pm100$&	$0.17\pm 0.08$&		$4.30\pm0.10$&	$9.0\pm0.5$&	Gonzalez 1997a, 1998 \\
6125&		0.06&			3.98&		&		Gratton et al.~1996 \\
6187&		-0.02&			4.13&		&		Marsakov et al.~1995 \\
6205&		&			&		&		Blackwell et al.~1994 \\
6212&		$0.09$&			4.17&		&		Edvardsson et al.~1993b \\
6198&		-0.03&			4.22&		&		Balachandran 1990 \\
6050&		0.06&			4.0&		&		Boesgaard \& Lavery 1986 \\
&		&			&		$9.0\pm0.4$&	Gray 1986 \\
6146&		-0.20&			4.60&		&		Thevenin \& Foy 1983 \\
6000&		-0.23&			3.91&		&		Hearnshaw 1974 \\
6072&		-0.14&			4.10&		&		Spite \& Spite 1973 \\
6072&		-0.11&			&		&		Powell 1970 \\ \tableline

\enddata
\normalsize
\end{deluxetable}

%\documentstyle[10pt,aaspp4]{article}
%\begin{document}
\begin{deluxetable}{cccccccccccccccc}
\scriptsize
\tablecolumns{16}
\tablewidth{0pc}
\tablecaption{Stellar Models for $\upsilon$ And}
\tablehead{ \colhead{$M_{*}$} & \colhead{ $\left[\frac{Fe}{H}\right]$ } & \colhead{ $T_{\rm eff}$ } & \colhead{ $L_{*}$} & \colhead{ Z } & \colhead{ X } & \colhead{ Y } & \colhead{ Age } & \colhead{ $M_{ce}$ } & \colhead{ PSH } & \colhead{ $R_{ce}$ } & \colhead{ $L_{*}$ } & \colhead{ $R_{*}$ } & \colhead{ $T_{\rm eff}$ } & \colhead{ $\log g$ } & \colhead{ $\frac{\delta Y}{\delta Z}$ } \nl 
\colhead{Solar} & \colhead{Solar} & \colhead{ K } & \colhead{Solar} & \colhead{ } & \colhead{ } & \colhead{ } & \colhead{Gyr} & \colhead{Solar} & \colhead{Solar} & \colhead{Solar} & \colhead{Solar} & \colhead{Solar} & \colhead{K} & \colhead{cgs} & \colhead{Solar}  } 
\startdata
1.359&	0.17&	6250&	3.333&	0.028&	0.689&	0.283&	2.5&	0.00204&	0.0841&	1.304&	3.336&	1.561&	6252&	4.18&	0.000\nl 
1.347&	0.17&	6250&	3.333&	0.028&	0.676&	0.297&	2.5&	0.00224&	0.0856&	1.297&	3.334&	1.561&	6250&	4.18&	1.500\nl 
1.339&	0.17&	6250&	3.333&	0.028&	0.667&	0.306&	2.6&	0.00238&	0.0866&	1.293&	3.332&	1.561&	6250&	4.18&	2.500\nl 
1.331&	0.17&	6250&	3.333&	0.028&	0.658&	0.315&	2.6&	0.00251&	0.0876&	1.289&	3.330&	1.561&	6249&	4.17&	3.500\nl 
\tableline
1.353&	0.17&	6150&	3.333&	0.028&	0.689&	0.283&	2.9&	0.00376&	0.0964&	1.310&	3.335&	1.613&	6150&	4.15&	0.000\nl 
1.339&	0.17&	6150&	3.333&	0.028&	0.676&	0.297&	3.0&	0.00406&	0.0979&	1.303&	3.332&	1.613&	6149&	4.15&	1.500\nl 
1.330&	0.17&	6150&	3.333&	0.028&	0.667&	0.306&	3.0&	0.00426&	0.0989&	1.299&	3.330&	1.612&	6148&	4.15&	2.500\nl 
1.320&	0.17&	6150&	3.333&	0.028&	0.658&	0.315&	3.1&	0.00445&	0.0998&	1.294&	3.327&	1.612&	6147&	4.14&	3.500\nl 
1.370&	0.17&	6350&	3.333&	0.028&	0.689&	0.283&	2.0&	0.00000&	0.0000&	1.291&	3.340&	1.513&	6353&	4.22&	0.000\nl 
1.357&	0.17&	6350&	3.333&	0.028&	0.676&	0.297&	2.1&	0.00000&	0.0000&	1.286&	3.337&	1.513&	6352&	4.21&	1.500\nl 
1.349&	0.17&	6350&	3.333&	0.028&	0.667&	0.306&	2.1&	0.00003&	0.0012&	1.283&	3.336&	1.513&	6351&	4.21&	2.500\nl 
1.341&	0.17&	6350&	3.333&	0.028&	0.658&	0.315&	2.1&	0.00031&	0.0144&	1.280&	3.334&	1.513&	6350&	4.21&	3.500\nl 
\tableline
1.315&	0.09&	6250&	3.333&	0.023&	0.694&	0.283&	2.9&	0.00241&	0.0871&	1.292&	3.330&	1.561&	6248&	4.17&	0.000\nl 
1.310&	0.09&	6250&	3.333&	0.023&	0.687&	0.289&	2.9&	0.00248&	0.0878&	1.290&	3.330&	1.561&	6248&	4.17&	1.500\nl 
1.306&	0.09&	6250&	3.333&	0.023&	0.683&	0.294&	2.9&	0.00253&	0.0882&	1.288&	3.330&	1.561&	6248&	4.17&	2.500\nl 
1.303&	0.09&	6250&	3.333&	0.023&	0.679&	0.298&	2.9&	0.00258&	0.0886&	1.286&	3.329&	1.561&	6248&	4.17&	3.500\nl 
1.403&	0.25&	6250&	3.333&	0.033&	0.684&	0.283&	2.2&	0.00183&	0.0815&	1.313&	3.335&	1.561&	6251&	4.20&	0.000\nl 
1.384&	0.25&	6250&	3.333&	0.033&	0.662&	0.305&	2.2&	0.00207&	0.0837&	1.305&	3.336&	1.561&	6251&	4.19&	1.500\nl 
1.371&	0.25&	6250&	3.333&	0.033&	0.647&	0.320&	2.3&	0.00223&	0.0851&	1.299&	3.337&	1.561&	6251&	4.19&	2.500\nl 
1.358&	0.25&	6250&	3.333&	0.033&	0.632&	0.334&	2.3&	0.00240&	0.0865&	1.293&	3.337&	1.561&	6251&	4.18&	3.500\nl 
\tableline
1.340&	0.17&	6250&	3.122&	0.028&	0.689&	0.283&	2.5&	0.00218&	0.0821&	1.259&	3.127&	1.511&	6252&	4.21&	0.000\nl 
1.329&	0.17&	6250&	3.122&	0.028&	0.676&	0.297&	2.5&	0.00237&	0.0835&	1.254&	3.125&	1.511&	6251&	4.20&	1.500\nl 
1.321&	0.17&	6250&	3.122&	0.028&	0.667&	0.306&	2.6&	0.00249&	0.0845&	1.250&	3.124&	1.511&	6251&	4.20&	2.500\nl 
1.313&	0.17&	6250&	3.122&	0.028&	0.658&	0.315&	2.6&	0.00261&	0.0854&	1.246&	3.123&	1.511&	6250&	4.20&	3.500\nl 
1.378&	0.17&	6250&	3.558&	0.028&	0.689&	0.283&	2.5&	0.00199&	0.0861&	1.350&	3.563&	1.613&	6252&	4.16&	0.000\nl 
1.367&	0.17&	6250&	3.558&	0.028&	0.676&	0.297&	2.5&	0.00214&	0.0875&	1.344&	3.561&	1.613&	6251&	4.16&	1.500\nl 
1.359&	0.17&	6250&	3.558&	0.028&	0.667&	0.306&	2.5&	0.00224&	0.0885&	1.340&	3.560&	1.613&	6251&	4.16&	2.500\nl 
1.351&	0.17&	6250&	3.558&	0.028&	0.658&	0.315&	2.6&	0.00235&	0.0895&	1.337&	3.559&	1.613&	6250&	4.15&	3.500\nl 
\enddata
\normalsize
\end{deluxetable}
%\end{document}

\begin{deluxetable}{ccccl}
\small
\tablecaption{Summary of Observational Data for $\rho^1$ Cnc}
\tablecolumns{5}
\tablewidth{0pc}
\tablehead{ \colhead{$T_{\rm eff}$ (K)}& \colhead{[Fe/H] (solar)}& \colhead{$\log g$ (cgs)}& \colhead{$v \sin i$ (km s$^{-1}$)}& \colhead{Reference} }
\startdata

$5336\pm90$&	$0.40\pm0.07$&		$4.47\pm0.10$&	$2.5\pm1.0$&	Fuhrmann et al.~1998 \\
$5250\pm70$&	$0.45\pm 0.03$&		$4.40\pm0.15$&	&		Gonzalez 1998a \\
$5150\pm75$&	$0.29\pm 0.06$&		$4.15\pm0.05$&	$1.4\pm0.5$&	Gonzalez 1997a \\
&		&			&		$1\pm1$& 	Baliunas et al.~1997 \\
&		$0.414\pm 0.096$&	&		&		Taylor 1991\tablenotemark{d} \\
$5100\pm150$&	0.20&			&		&		Arribas \& Martinez-Roger 1989 \\
5350&		0.44&			&		&		Campbell 1978 \\
$5200$&		$0.24\pm 0.09$&		$4.5$&		&		Perrin 1977 \\
5140&	   	0.30&			4.4& 		&		Oinas 1977 \\
5196& 		0.11&			4.4&		&		Oinas 1974 \\
4460&		-0.15&			&		&		Bakos 1971 \\ \tableline
\enddata
\tablenotetext{d}{This value is based on a statistical analysis of
previous measurements and not any new observations.}
\normalsize
\end{deluxetable}

%\documentstyle[10pt,aaspp4]{article}
%\begin{document}
\begin{deluxetable}{cccccccccccccccc}
\scriptsize
\tablecolumns{16}
\tablewidth{0pc}
\tablecaption{Stellar Models for $\rho^1$ Cnc}
\tablehead{ \colhead{$M_{*}$} & \colhead{ $\left[\frac{Fe}{H}\right]$ } & \colhead{ $T_{\rm eff}$ } & \colhead{ $L_{*}$} & \colhead{ Z } & \colhead{ X } & \colhead{ Y } & \colhead{ Age } & \colhead{ $M_{ce}$ } & \colhead{ PSH } & \colhead{ $R_{ce}$ } & \colhead{ $L_{*}$ } & \colhead{ $R_{*}$ } & \colhead{ $T_{\rm eff}$ } & \colhead{ $\log g$ } & \colhead{ $\frac{\delta Y}{\delta Z}$ } \nl 
\colhead{Solar} & \colhead{Solar} & \colhead{ K } & \colhead{Solar} & \colhead{ } & \colhead{ } & \colhead{ } & \colhead{Gyr} & \colhead{Solar} & \colhead{Solar} & \colhead{Solar} & \colhead{Solar} & \colhead{Solar} & \colhead{K} & \colhead{cgs} & \colhead{Solar}  } 
\startdata
1.029&	0.45&	5300&	0.608&	0.053&	0.613&	0.334&	2.4&	0.04444&	0.0798&	0.647&	0.612&	0.927&	5308&	4.52&	1.500\nl 
0.951&	0.45&	5300&	0.608&	0.053&	0.579&	0.368&	8.4&	0.04583&	0.0825&	0.628&	0.610&	0.927&	5305&	4.48&	2.500\nl 
0.873&	0.45&	5300&	0.608&	0.053&	0.544&	0.403&	14.3&	0.04722&	0.0853&	0.609&	0.609&	0.927&	5303&	4.45&	3.500\nl 
\tableline
1.048&	0.39&	5275&	0.608&	0.046&	0.671&	0.283&	1.7&	0.04703&	0.0809&	0.656&	0.606&	0.936&	5270&	4.52&	0.000\nl 
0.971&	0.39&	5275&	0.608&	0.046&	0.630&	0.324&	7.9&	0.04798&	0.0835&	0.637&	0.608&	0.936&	5275&	4.48&	1.500\nl 
0.920&	0.39&	5275&	0.608&	0.046&	0.602&	0.351&	12.1&	0.04862&	0.0851&	0.624&	0.610&	0.936&	5279&	4.46&	2.500\nl 
0.868&	0.39&	5275&	0.608&	0.046&	0.575&	0.379&	16.3&	0.04925&	0.0868&	0.611&	0.611&	0.936&	5282&	4.43&	3.500\nl 
0.978&	0.45&	5325&	0.608&	0.053&	0.579&	0.368&	5.4&	0.04404&	0.0803&	0.631&	0.610&	0.919&	5329&	4.50&	2.500\nl 
0.896&	0.45&	5325&	0.608&	0.053&	0.544&	0.403&	11.6&	0.04466&	0.0829&	0.613&	0.611&	0.919&	5330&	4.47&	3.500\nl 
\tableline
1.006&	0.42&	5300&	0.608&	0.049&	0.622&	0.329&	4.1&	0.04541&	0.0808&	0.642&	0.609&	0.927&	5302&	4.51&	1.500\nl 
0.945&	0.42&	5300&	0.608&	0.049&	0.591&	0.360&	8.9&	0.04620&	0.0828&	0.627&	0.609&	0.927&	5303&	4.48&	2.500\nl 
0.885&	0.42&	5300&	0.608&	0.049&	0.560&	0.390&	13.7&	0.04699&	0.0847&	0.612&	0.609&	0.927&	5303&	4.45&	3.500\nl 
1.045&	0.48&	5300&	0.608&	0.057&	0.603&	0.340&	1.2&	0.04348&	0.0792&	0.651&	0.616&	0.928&	5315&	4.52&	1.500\nl 
0.959&	0.48&	5300&	0.608&	0.057&	0.565&	0.378&	7.7&	0.04580&	0.0823&	0.629&	0.609&	0.927&	5303&	4.48&	2.500\nl 
0.872&	0.48&	5300&	0.608&	0.057&	0.527&	0.416&	14.2&	0.04811&	0.0854&	0.607&	0.603&	0.927&	5290&	4.45&	3.500\nl 
\tableline
0.991&	0.45&	5300&	0.569&	0.053&	0.579&	0.368&	2.8&	0.04437&	0.0775&	0.622&	0.573&	0.897&	5308&	4.53&	2.500\nl 
0.898&	0.45&	5300&	0.569&	0.053&	0.544&	0.403&	10.1&	0.04585&	0.0807&	0.601&	0.568&	0.897&	5298&	4.49&	3.500\nl 
1.020&	0.45&	5325&	0.650&	0.053&	0.613&	0.334&	4.7&	0.04403&	0.0826&	0.657&	0.653&	0.950&	5330&	4.49&	1.500\nl 
0.950&	0.45&	5325&	0.650&	0.053&	0.579&	0.368&	9.7&	0.04559&	0.0850&	0.640&	0.650&	0.950&	5325&	4.46&	2.500\nl 
0.881&	0.45&	5325&	0.650&	0.053&	0.544&	0.403&	14.7&	0.04714&	0.0874&	0.622&	0.648&	0.950&	5320&	4.43&	3.500\nl 
\enddata
\normalsize
\end{deluxetable}
%\end{document}

\begin{deluxetable}{ccccl}
\small
\tablecaption{Summary of Observational Data for $\rho$ CrB}
\tablecolumns{5}
\tablewidth{0pc}
\tablehead{ \colhead{$T_{\rm eff}$ (K)}& \colhead{[Fe/H] (solar)}& \colhead{$\log g$ (cgs)}& \colhead{$v \sin i$ (km s$^{-1}$)}& \colhead{Reference} }
\startdata

$5821\pm80$&	$-0.24\pm0.08$&	$4.12\pm0.10$&	$1.0\pm1.0$&	Fuhrmann et al.~1998 \\
$5750\pm 75$&	$-0.29\pm 0.06$&$4.10\pm0.05$&	$\sim1.5$&	Gonzalez 1998 \\
&		&		4.27&		&		Kunzl et al.~1997 \\
5745&		-0.22&		4.11&		&		Gratton et al.~1996 \\
5905&	 	-0.24&		4.20&		&		Marsakov et al.~1995 \\
5782&		-0.26&		4.24&		&		Edvardsson et al.~1993b \\ 
&		&		&		$1.5\pm1.0$&	Soderblom 1982 \\
5780&		-0.17&		3.98&		&		Hearnshaw 1974 \\ 
5860&		-0.14&		&		&		Alexander 1967 \\
5663&		-0.20&		&		&		Wallerstein 1962 \\ \tableline
\enddata
\normalsize
\end{deluxetable}

%\documentstyle[10pt,aaspp4]{article}
%\begin{document}
\begin{deluxetable}{cccccccccccccccc}
\scriptsize
\tablecolumns{16}
\tablewidth{0pc}
\tablecaption{Stellar Models for $\rho$ CrB}
\tablehead{ \colhead{$M_{*}$} & \colhead{ $\left[\frac{Fe}{H}\right]$ } & \colhead{ $T_{\rm eff}$ } & \colhead{ $L_{*}$} & \colhead{ Z } & \colhead{ X } & \colhead{ Y } & \colhead{ Age } & \colhead{ $M_{ce}$ } & \colhead{ PSH } & \colhead{ $R_{ce}$ } & \colhead{ $L_{*}$ } & \colhead{ $R_{*}$ } & \colhead{ $T_{\rm eff}$ } & \colhead{ $\log g$ } & \colhead{ $\frac{\delta Y}{\delta Z}$ } \nl 
\colhead{Solar} & \colhead{Solar} & \colhead{ K } & \colhead{Solar} & \colhead{ } & \colhead{ } & \colhead{ } & \colhead{Gyr} & \colhead{Solar} & \colhead{Solar} & \colhead{Solar} & \colhead{Solar} & \colhead{Solar} & \colhead{K} & \colhead{cgs} & \colhead{Solar}  } 
\startdata
0.874&	-0.23&	5750&	1.796&	0.011&	0.706&	0.283&	14.6&	0.03400&	0.1256&	0.873&	1.796&	1.354&	5750&	4.12&	0.000\nl 
0.886&	-0.23&	5750&	1.796&	0.011&	0.718&	0.271&	14.3&	0.03341&	0.1250&	0.879&	1.796&	1.354&	5750&	4.12&	1.500\nl 
0.894&	-0.23&	5750&	1.796&	0.011&	0.725&	0.264&	14.1&	0.03301&	0.1247&	0.883&	1.796&	1.354&	5750&	4.13&	2.500\nl 
0.902&	-0.23&	5750&	1.796&	0.011&	0.733&	0.256&	14.0&	0.03262&	0.1243&	0.887&	1.795&	1.354&	5750&	4.13&	3.500\nl 
\tableline
0.863&	-0.23&	5675&	1.796&	0.011&	0.706&	0.283&	15.5&	0.04391&	0.1334&	0.865&	1.795&	1.390&	5675&	4.09&	0.000\nl 
0.875&	-0.23&	5675&	1.796&	0.011&	0.718&	0.271&	15.3&	0.04330&	0.1330&	0.871&	1.795&	1.390&	5674&	4.09&	1.500\nl 
0.883&	-0.23&	5675&	1.796&	0.011&	0.725&	0.264&	15.1&	0.04290&	0.1326&	0.875&	1.794&	1.390&	5674&	4.10&	2.500\nl 
0.890&	-0.23&	5675&	1.796&	0.011&	0.733&	0.256&	15.0&	0.04249&	0.1323&	0.879&	1.794&	1.390&	5673&	4.10&	3.500\nl 
0.888&	-0.23&	5825&	1.796&	0.011&	0.706&	0.283&	13.4&	0.02581&	0.1177&	0.882&	1.797&	1.319&	5825&	4.15&	0.000\nl 
0.900&	-0.23&	5825&	1.796&	0.011&	0.718&	0.271&	13.2&	0.02536&	0.1171&	0.887&	1.796&	1.319&	5825&	4.15&	1.500\nl 
0.909&	-0.23&	5825&	1.796&	0.011&	0.725&	0.264&	13.0&	0.02507&	0.1167&	0.891&	1.796&	1.319&	5825&	4.16&	2.500\nl 
0.917&	-0.23&	5825&	1.796&	0.011&	0.733&	0.256&	12.8&	0.02477&	0.1163&	0.895&	1.796&	1.319&	5825&	4.16&	3.500\nl 
\tableline
0.896&	-0.17&	5750&	1.796&	0.013&	0.704&	0.283&	13.8&	0.03231&	0.1242&	0.884&	1.796&	1.354&	5750&	4.13&	0.000\nl 
0.906&	-0.17&	5750&	1.796&	0.013&	0.713&	0.274&	13.6&	0.03191&	0.1237&	0.888&	1.796&	1.354&	5750&	4.13&	1.500\nl 
0.912&	-0.17&	5750&	1.796&	0.013&	0.720&	0.268&	13.5&	0.03163&	0.1234&	0.891&	1.796&	1.354&	5750&	4.13&	2.500\nl 
0.919&	-0.17&	5750&	1.796&	0.013&	0.726&	0.262&	13.4&	0.03136&	0.1231&	0.894&	1.796&	1.354&	5750&	4.14&	3.500\nl 
\tableline
0.862&	-0.23&	5750&	1.673&	0.011&	0.706&	0.283&	15.3&	0.03306&	0.1208&	0.846&	1.673&	1.307&	5750&	4.14&	0.000\nl 
0.875&	-0.23&	5750&	1.673&	0.011&	0.718&	0.271&	15.0&	0.03258&	0.1203&	0.852&	1.673&	1.307&	5750&	4.15&	1.500\nl 
0.883&	-0.23&	5750&	1.673&	0.011&	0.725&	0.264&	14.8&	0.03226&	0.1200&	0.855&	1.673&	1.307&	5750&	4.15&	2.500\nl 
0.891&	-0.23&	5750&	1.673&	0.011&	0.733&	0.256&	14.6&	0.03194&	0.1197&	0.859&	1.673&	1.307&	5750&	4.15&	3.500\nl 
0.887&	-0.23&	5750&	1.928&	0.011&	0.706&	0.283&	13.8&	0.03464&	0.1303&	0.903&	1.928&	1.403&	5750&	4.09&	0.000\nl 
0.899&	-0.23&	5750&	1.928&	0.011&	0.718&	0.271&	13.6&	0.03403&	0.1297&	0.909&	1.928&	1.403&	5750&	4.10&	1.500\nl 
0.907&	-0.23&	5750&	1.928&	0.011&	0.725&	0.264&	13.4&	0.03362&	0.1293&	0.913&	1.928&	1.403&	5750&	4.10&	2.500\nl 
0.915&	-0.23&	5750&	1.928&	0.011&	0.733&	0.256&	13.3&	0.03321&	0.1289&	0.917&	1.928&	1.403&	5750&	4.11&	3.500\nl 
\enddata
\normalsize
\end{deluxetable}
%\end{document}

\end{document}